# Kinetics of grain-boundary nucleated transformations in rectangular geometries and one paradox relating to Cahn's model

Nikolay V. Alekseechkin

Akhiezer Institute for Theoretical Physics, National Science Centre "Kharkov Institute of Physics and Technology", Akademicheskaya Street 1, Kharkov 61108, Ukraine
Email: n.alex@kipt.kharkov.ua

It was found by Jägle and Mittenmeijer [Acta Mater. 59 (2011) 5775-5786] as a result of computer simulations that Cahn's equation underestimates pronouncedly the transformed fraction in reality. To test the hypothesis that a strong correlation in the arrangement of grains in the Cahn model (they are "tied" to planes) can be responsible for this phenomenon, the paradox of packing is described: the structures composed of the same elements (random parallelepipeds) transforming in the same way, but packed differently, give different transformation rates. One of these structures is a special Cahn model – the grain structure of parallelepipeds formed by three sets of random parallel planes orthogonal to each other. An alternative structure consists of the same parallelepipeds, but randomly packed; it looks like a real grain structure. It is shown analytically that the kinetics of transformation in the first structure is considerably slowed down in comparison with that in the second. For the various intermediate structures under consideration, the kinetic curves are between these two, so that the following rule is established: the transformation kinetics is accelerated, when the correlation in the arrangement of the parallelepipeds weakens. Preliminarily, volume-fraction expressions are obtained for the systems of an infinite number of parallel planes arranged both regularly and randomly. As a special case of random arrangement, a non-Poissonian point process (the second-order Erlang process) of arrangement of planes is considered for the first time. The exact volume-fraction expression obtained for this case shows that it cannot be derived by Cahn's method, i.e. the extended-volume approach is applicable only to Poisson processes. The volume fraction equations for regular planes are used to study cubic grain structures, both regular and random. It is shown that Cahn's equation underestimates the transformation kinetics in both regular and random structures with four different size distributions of cubes; the degree of underestimation depends on the size distribution, being the largest in the regular structure.

## 1. Introduction

The classical Kolmogorov-Johnson-Mehl-Avrami (KJMA) theory [1-5] solves the problem of calculating the volume fraction (VF) of a new phase in the process of crystallization under certain model assumptions. In view of the importance of this theory and its broad applications, the verification of underlying model assumptions and various extensions of the original theory, including extensions beyond its limitations, continued in recent decades [6-17]. While the VF is an integral quantity, a more detailed study of the transformation process and the resulting structure is performed using Secimoto's [18] $m$-point correlation functions [19-23], microstructural descriptors [24-29], grain-size distribution functions [30-35], and other characteristics [36, 37]. Double-logarithmic VF plots, together with the corresponding temporal behavior of the Avrami exponent, provide information on the mechanisms of nucleation and growth.

One of the extensions of the KJMA theory is calculating the VF of a new phase nucleated at lower-dimension objects such as surfaces, lines, and points. Originally, this problem was first considered by Cahn [38] as applied to nucleation at faces, edges, and corners of grains in the framework of Avrami's extended-volume concept [4, 5]; later, this model was revisited and generalized [7, 39-41]. It should be noted that VF calculation for progressive nucleation at points was done by Avrami himself [4, 5] who assumed that nucleation generally occurs by this mechanism. Some of the problems mentioned above, specifically, grain-size distribution and grain-boundary nucleated transformation were already considered by Johnson and Mehl [2]. In the latter case, they represented grains as equal spheres, and also assumed grains "isolated" from each other when a growing nucleus could not cross the grain boundaries. Later, Johnson and Mehl's model was supplanted by Cahn's one.

The model of spherical grains has recently been revisited and considerably improved [42]: firstly, the grains are not isolated from each other and are distributed in size; secondly, the rigorous *critical-region* (CR) method [6-8] which is a development of Kolmogorov's one [1] was applied to obtain the VF expression for this system. The transformation kinetics (TK) is determined here (in addition to the distribution function) by the characteristic parameter $\overline{\alpha}_s$ which is the mean number of nuclei born on the boundary of the grain of mean size $\overline{R}_0$ for the time $t^* = \overline{R}_0 / u$, where $u$ is the nucleus growth velocity. It was shown that the TK in this model qualitatively differs from that of Cahn's model. At large $\overline{\alpha}_s$ values, the double-logarithmic VF plot deviates from Cahn's one and ends with a steep bend which corresponds to a sharp increase in the Avrami exponent $n$ at the late stage, whereas $n \to 1$ in the Cahn model. This bend exists in two opposite cases - a very narrow ($\delta$-shaped) and a wide grain size distribution. It was shown by computer simulations [30, 35] and analytical calculations [2, 42] that a process with continuous nucleation results just in a wide grain size distribution. The similar form of double-logarithmic VF plots obtained by experimentalists for the crystallization of bulk metallic glasses [43] suggests that (i) these glasses have a grain structure (as opposed to the concept of homogeneous "frozen" liquid) and nucleation occurs at grain boundaries; (ii) this structure was formed as a result of continuous nucleation and growth of non-crystalline nuclei (clusters) [44, 45]. In this way, the result of studying TK allows us to make important conclusions about the structure of matter and the way of its formation.

The different behavior of the Avrami exponents mentioned above is reflected in the behavior of the corresponding VFs. It was established for the first time by Jägle and Mittenmeijer [46] that Cahn's model underestimates pronouncedly the TK in reality. The present report aims to investigate in detail the cause of this phenomenon. The following two cases should be distinguished. (i) Comparison of the TK in any structure of *identical* elements of size $R_0$ (e.g., spheres) with the corresponding (i.e. containing the same $\omega$, the area of boundaries in unit volume) Cahn expression. (ii) Comparison of the TK in a structure of *size-distributed* elements with the corresponding Cahn expression. As grains are size-distributed in the Cahn model, the underestimation in the first case is easily explained by the contribution of large grains with $R > R_0$; just these grains delay the transformation process, "stretching" it in time. Cahn's VF has the exponential tail $X_C(t) \to 1 - \exp(-2\omega ut)$ originating from random planes and corresponding to $n = 1$, whereas the cubic-power asymptotics $X \to 1 - (1 - ut / R_0)^3$ holds in the system of identical elements at sufficiently large $\alpha_s$ values, giving a sharp increase in $n$ [42]. This case is also illustrated here by comparing the TK in regular and random structures of both plates and planes.

Cahn's equation involves a single parameter $\omega$ which can be adjustable at a fixed $\overline{\alpha}_s$ value. However, the Avrami exponent does not depend on $\omega$; it is determined by the model itself and cannot be changed by a change in $\omega$ value. The latter only shifts the kinetic curve as a whole, so that it can intersect a comparing curve, underestimating it at the late stage due to the exponential tail [42]. So, the temporal behavior of the Avrami exponent in the model of spherical grains cannot be reproduced by any "corrected" Cahn equation in principle.

The second case is more complicated. First, it is unclear whether this underestimation is systematic (i. e. holds for all grain size distributions) or is absent for some distributions. Second, as two size-distributed structures are compared, the reason for the underestimation is not evident; individual grains in both structures are transformed in the same way. However, there is an essential distinction of the Cahn model from a real grain structure: the arrangement of grains in this model is strongly correlated - infinite arrays of grains are "tied" to their own planes. To understand how this correlation affects the TK and whether it is the cause of underestimation, it is necessary to compare Cahn's TK with that in a grain structure with the same size distribution as in the Cahn model. However, this distribution is unknown and its obtaining is not an easy task.

Therefore, a *particular* Cahn model is considered here – the grain structure of random parallelepipeds which is formed by three sets of random planes orthogonal to each other. Since the TK in this model is described by the same equation as in the general Cahn model, both these models are physically equivalent, if they use the same $\omega$-value. However, all calculations, e.g., obtaining the grain size distribution, in the particular model are much simpler. The same parallelepipeds but repacked randomly form the grain structure similar to a real one; it has the same volume distribution of grains, as the initial structure. Comparison of the TK in these two structures positively answers the question posed above: indeed, the correlation slows down the TK. Consideration of some intermediate structures between these two confirms this result – the TK accelerates with a weakening of the degree of correlation in the arrangement of parallelepipeds. This phenomenon is called here the paradox of packing: structures consisting of the same structural elements, but packed in different ways, give different rates of transformation.

Preliminarily, the systems of parallel planes both with regular and random arrangement are considered. The system of random planes with the Poisson process of their positions is a one-dimensional Cahn model. Alongside with this case, the case of a non-Poissonian process of the arrangement of planes is studied for the first time. The resulting exact VF expression shows that it cannot be obtained by the Cahn method, i.e. the extended-volume approach is not universal.

The paper is organized as follows. In Section 2, the process of surface-nucleated transformation of a single infinite plate is considered; both nonisothermal and isothermal equations are derived and analized. The TK in the ensemble of width-distributed plates is studied in Section 3. The results of these two Sections are employed in Section 4, where the systems of an innifinite number of parallel planes arranged either regularly or randomly are considered. The paradox of packing is described in details in Section 5. The TK in cubic structures, both regular and random, is studied in Section 6. Conclusions Section finalizes the paper.

## 2. Single plate

### 2.1. Model and nonisothermal equations

We consider the process of crystallization of an infinite plate of width $L = 2R_0$ under the nucleation of new-phase centers on its surface (the two planes) with the nucleation rate $I_s(t)$ and the growth velocity $u(t)$. In what follows, the same notations are used as in Ref. [42] for ease of comparison and to demonstrate the generality of the approach employed. Let us take at random a point $O'$ in the plate. Let it be at a distance $r$ from the middle plane (the point $O$), Fig. 1. We find the probability $Q(r,t)$ that the point $O'$ is untransformed at time $t$.

For this purpose, we specify the CR for the point $O'$ - the sphere (for the spherical shape of nuclei) of radius $R(t',t)$, where

$$R(t',t) = \int_{t'}^{t} u(\tau)d\tau$$

i.e. the CR radius at time $t'$ equals to the radius of a $t'$-age nucleus at time $t$. At time $t'$, the CR boundary moves at the velocity $u(t')$ towards the point $O'$, so that in the time interval $0 \le t' \le t$ the CR radius decreases from the largest value $R(0,t) \equiv R_m(t)$ to $R(t,t) \equiv 0$. In order for the point $O'$ to be untransformed at time $t$, it is necessary and sufficient that a new-phase centre does not appear inside the CR in the time interval $0 \le t' \le t$. The probability of this event is

$$Q(r,t) = \exp[-Y(r,t)] \tag{1}$$

which is a general equation for the probability of no nucleation event in the Poisson nucleation process.

As the centers of a new phase appear on the plate surface only, the surface part lying within the CR (Fig. 1) should be taken for calculating $Q(r,t)$; denoting its area by $\Omega(r;t',t)$, we have [42]

$$Y(r,t) = \int_{0}^{t} I_s(t')\Omega(r;t',t)dt' \tag{2}$$

Depending on $t$, $t'$, and $r$, the area $\Omega(r;t',t)$ can be equal to 0, $S_1(r;t',t)$, or $S_1(r;t',t) + S_2(r;t',t)$, where

$$S_1(r;t',t) = \pi\left[R^2(t',t) - (R_0 - r)^2\right] \tag{3a}$$

is the area of either the circle $a_1a_1$ or $a_3a_3$ in Fig. 1, and

$$S_2(r;t',t) = \pi\left[R^2(t',t) - (R_0 + r)^2\right] \tag{3b}$$

is the area of the circle $b_1b_1$.

The probability for the random point $O'$ to fall at the distances $[r, r+dr]$ either to the right or left of the point $O$ is $2dr/L = dr/R_0$; $Q(r,t)$ is the same for the right and left parts of the plate $(r > 0)$, in view of obvious symmetry relatively the middle plane. Hence, the probability for the point $O'$ to fall into the untransformed part of the plate is

$$Q^{(pt)}(t) = \frac{1}{R_0}\int_0^{R_0} Q(r,t)dr \tag{4}$$

Differently from the case of spherical domain, where the similar equation gives the ensemble-averaged VF value [42], here this is an exact VF value of the plate; as the plate is infinite, $Q(t)$ is a *self-averaging* quantity here [42], similarly to the case of KJMA model.

For the calculation of $Y(r,t)$, we determine times $t_1$ and $t_2$ by the equations

$$R_m(t_1) = R_0, \quad R_m(t_2) = 2R_0 \tag{5}$$

and have the following three cases with respect to time $t$, similarly to Refs. [7, 42].

(1) $R_m(t) < R_0$: $t < t_1$

We determine the distance $r_0 = R_0 - R_m(t)$ and the time $t_m(r,t)$ by the equation $R(t_m, t) = R_0 - r$ corresponding to CR boundary position 4 in Fig. 1. At $0 \le r \le r_0$, the CR does not include the plate surface, accordingly, $\Omega(r;t',t) = 0$. At $r_0 < r \le R_0$, the CR crosses the plane in the time interval $0 \le t' < t_m(r,t)$, accordingly, $\Omega(r;t',t) = S_1(r;t',t)$ (as in CR boundary position 3). Further, in the interval $t_m(r,t) \le t' \le t$ the CR is empty and $\Omega(r;t',t) = 0$. Summarizing,

$$Y_1(r,t) = \begin{cases} 0, & 0 \le r \le r_0 \\ \int_0^{t_m(r,t)} I_s(t')S_1(r;t',t)dt', & r_0 < r \le R_0 \end{cases} \tag{6a}$$

(2) $R_0 \le R_m(t) \le 2R_0$: $t_1 \le t \le t_2$

We determine the distance $r_0' = R_m(t) - R_0$ and the time $t_m'(r,t)$ by the equation $R(t_m', t) = R_0 + r$ corresponding to CR boundary position 2. Let us consider the case $0 \le r < r_0'$. In Fig. 1, the CR boundary positions are shown for this case at different times $t'$. In the time interval $0 \le t' \le t_m'(r,t)$, the CR crosses both the planes (position 1) and includes circles $b_1b_1$ and $a_1a_1$, accordingly, $\Omega(r;t',t) = S_2(r;t',t) + S_1(r;t',t)$. In the interval $t_m'(r,t) < t' \le t_m(r,t)$, we have the equality $\Omega(r;t',t) = S_1(r;t',t)$ corresponding to position 3. And in the remaining interval $t_m(r,t) < t' \le t$, the equality $\Omega(r;t',t) = 0$ holds. At $r_0' \le r \le R_0$, we have $\Omega(r;t',t) = S_1(r;t',t)$ in the interval $0 \le t' < t_m(r,t)$ and $\Omega(r;t',t) = 0$ for $t_m(r,t) \le t' \le t$. In this way,

$$Y_2(r,t) = \begin{cases} \int_0^{t_m'(r,t)} I_s(t')\left[S_2(r;t',t) + S_1(r;t',t)\right]dt' + \int_{t_m'(r,t)}^{t_m(r,t)} I_s(t')S_1(r;t',t)dt', & 0 \le r < r_0' \\ \int_0^{t_m(r,t)} I_s(t')S_1(r;t',t)dt', & r_0' \le r \le R_0 \end{cases} \quad (6b)$$

(3) $R_m(t) > 2R_0$: $t > t_2$

As follows from the foregoing, in this case

$$\Omega(r;t',t) = \begin{cases} S_2(r;t',t) + S_1(r;t',t), & 0 \le t' \le t_m'(r,t) \\ S_1(r;t',t), & t_m'(r,t) < t' < t_m(r,t) \\ 0, & t_m(r,t) \le t' \le t \end{cases} \quad (6c)$$

for arbitrary $r$-value. Accordingly,

$$Y_3(r,t) = \int_0^{t_m'(r,t)} I_s(t')\left[S_2(r;t',t) + S_1(r;t',t)\right]dt' + \int_{t_m'(r,t)}^{t_m(r,t)} I_s(t')S_1(r;t',t)dt', \quad 0 \le r \le R_0 \quad (6d)$$

The VF $Q_i(t)$ in each case is calculated according to Eqs. (1) and (4); e.g., we have for case (1):

$$Q_1^{(pt)}(t) = \frac{1}{R_0}\left\{\int_0^{r_0} dr + \int_{r_0}^{R_0} \exp\left[-\int_0^{t_m(r,t)} I_s(t')S_1(r;t',t)dt'\right]dr\right\}$$

$$= \left(1 - \frac{R_m(t)}{R_0}\right) + \frac{1}{R_0}\int_{R_0 - R_m(t)}^{R_0} \exp\left[-\int_0^{t_m(r,t)} I_s(t')S_1(r;t',t)dt'\right]dr, \quad t < t_1 \quad (7)$$

Accordingly, the transformed VF is $X_i^{(pt)}(t) = 1 - Q_i^{(pt)}(t)$.

## 2.2. Isothermal equations

In the case of isothermal crystallization (constant $I_s$ and $u$), the key quantities defined above acquire the following form:

$$R(t',t) = u(t - t'), \quad R_m(t) = ut, \quad r_0 = R_0 - ut, \quad t_m(r,t) = t - \frac{R_0 - r}{u}, \quad r_0' = ut - R_0,$$

$$t_m'(r,t) = t - \frac{R_0 + r}{u} \quad (8)$$

We introduce the dimensionless time $\tau = ut/R_0 = 2ut/L$ and distance $x = r/R_0$ as well as the characteristic parameter $\alpha_s = (\pi/3)(I_s/u)R_0^3$. After calculating the integrals in the above expressions, we obtain the following equations:

(1) $\tau < 1$:

$$Q_1^{(pt)}(\tau) = (1-\tau) + \int_{1-\tau}^{1} \mathrm{e}^{-\alpha_s \varphi_1(x,\tau)}\, dx\,, \quad \varphi_1(x,\tau) = \tau^3 - 3\tau(1-x)^2 + 2(1-x)^3 \tag{9a}$$

(2) $1 \le \tau \le 2$:

$$Q_2^{(pt)}(\tau) = \int_{0}^{\tau-1} \mathrm{e}^{-\alpha_s \varphi_2(x,\tau)}\, dx + \int_{\tau-1}^{1} \mathrm{e}^{-\alpha_s \varphi_1(x,\tau)}\, dx\,, \quad \varphi_2(x,\tau) = 2\left[\tau^3 - 3\tau(1+x^2) + 6x^2 + 2\right] \tag{9b}$$

(3) $\tau > 2$:

$$Q_3^{(pt)}(\tau) = \int_{0}^{1} \mathrm{e}^{-\alpha_s \varphi_2(x,\tau)}\, dx \tag{9c}$$

It should be noted that $\varphi_1(x,\tau) = 0$ at $x = 1-\tau$ and $\varphi_1(x,\tau) = \tau^3$ at $x = 1$. These equations can be easily rewritten for the VF $X^{(pt)}(\tau)$; e.g., Eq. (9a) goes into

$$X_1^{(pt)}(\tau) = \int_{1-\tau}^{1} \left[1 - \mathrm{e}^{-\alpha_s \varphi_1(x,\tau)}\right] dx\,, \quad \tau < 1 \tag{9d}$$

Isothermal equations for $X^{(pt)}(t)$ were also obtained in Ref. [17] by the time-cone [47, 48] method; the results of numerical simulation of film transformation are reported in Ref. [49].

It is also of interest to get an equation for the untransformed fraction $Q_s(\tau)$ of the surface itself. It is given by Eq. (1) taken at $r = R_0$; according to Eqs. (6a-d), the functions $\varphi_1(x,\tau)$ and $\varphi_2(x,\tau)$ are taken at $x = 1$. Doing so, we get

$$Q_s(\tau) = \begin{cases} \mathrm{e}^{-\alpha_s \tau^3}, & 0 < \tau \le 2 \\ \mathrm{e}^{-2\alpha_s(\tau^3 - 6\tau + 8)}, & \tau > 2 \end{cases} \tag{10}$$

It is seen that for $\tau < 2$ the law of plate-surface transformation is the same as the law of transformation of a single plane, as it must, whereas the case of $\tau > 2$ reflects the influence of another plane: at time $\tau > 2$ ($ut > L$), a nucleus appeared on one plane can cross another plane. So, the parameter $\alpha_s$ characterizes the rate of surface transformation; the characteristic time of surface transformation is $\tau_s \sim \alpha_s^{-1/3}$.

Defining by $t^* = R_0/u$ the time corresponding to $\tau = 1$, the following representation for $\alpha_s$ is obtained:

$$\alpha_s = \frac{\pi}{3} I_s t^* R_0^2 \tag{11}$$

Hence, this parameter is proportional to the mean number of nuclei formed on the area $R_0^2$ during time $t^*$.

Having VF equations, we can find the Avrami-exponent temporal behavior for the plate *isothermal* transformation:

$$n^{(pt)}(\tau) = \frac{d\ln(-\ln Q^{(pt)}(\tau))}{d\ln\tau} \tag{12}$$

Later, the *boundary saturation* (BS) limit $\alpha_s \to \infty$ will be essentially employed. In this limit, transformation proceeds by the 1D growth of the film formed on a plane at $t' = 0$. Real kinetic curves approach the BS-limit curve with increasing $\alpha_s$ s, as the BS phenomenon [38] at a finite $\alpha_s$ value occurs at the characteristic time $\tau_s$; the change in the transformation mode results in the corresponding change in the Avrami exponent [38]. The BS-limit VF is

$$Q_{BS}^{(pt)}(\tau) = \begin{cases} 1-\tau, & \tau \le 1 \\ 0, & \tau > 1 \end{cases} \tag{13}$$

The corresponding Avrami exponent is

$$n_{BS}(\tau) = \tau \frac{d\ln(-\ln(1-\tau))}{d\tau} = -\frac{\tau}{(1-\tau)\ln(1-\tau)} \tag{14}$$

In Fig. 2a, the plots of $X^{(pt)}(\tau)$ are shown for different $\alpha_s$ values; the functions $X_s(\tau) = 1 - Q_s(\tau)$ are also plotted for comparison. The kinetics of surface transformation is close to that of plate transformation at small $\alpha_s$ s, whereas the former outstrips the latter at $\alpha_s$>1; this effect is pronounced at large $\alpha_s$ s. In this way, while the plate and its surface complete transformation at about the same time at small $\alpha_s$ s, the surface becomes transformed much earlier than the plate at large $\alpha_s$ s; this is just the BS phenomenon occuring at time $\tau_s$. It is seen that the BS-limit line $X_{BS}^{(pt)}(\tau) = \tau$ for $\tau \le 1$ and $X_{BS}^{(pt)}(\tau) = 1$ for $\tau > 1$ is asymptotic for the kinetic curves $X^{(pt)}(\tau)$ approaching it at large $\alpha_s$ s.

Although the Avrami exponent in Fig. 2b generally varies in time from 4 to 3, its behavior is quite different at small and large $\alpha_s$ s; it is smooth in the first case and “irregular” in the second, similarly to the spherical particle transformation [42]. The following expansion holds at $\tau \to 0$:

$$\int_{1-\tau}^{1} \mathrm{e}^{-\alpha_s \varphi_1(x,\tau)}\, dx = \tau - \frac{\alpha_s}{2}\tau^4 + \ldots \tag{15}$$

Thus,

$$Q_1(\tau) = 1 - \frac{1}{2}\alpha_s \tau^4\,, \quad X_1(\tau) = \frac{1}{2}\alpha_s \tau^4 = \frac{1}{2}Y_b\,, \quad Y_b \equiv \alpha_s \tau^4 = \frac{\pi}{3} I_b u^3 t^4\,, \quad I_b = \omega_0 I_s\,,$$

$$\omega_0 = \frac{2}{L} = \frac{1}{R_0} \tag{16}$$

where $\omega_0$ is the mean area of plate surface in unit volume. This equation can be regarded as the first term of expansion of the exponential $Q_1(\tau) = \exp(-(1/2)Y_b(\tau))$; hence, $n = 4$ in this limit. $Y_b$ is the "extended volume" in the Avrami theory [4, 5]; the multiplier 1/2 reflects the fact that only half of the sphere born on the surface grows into the plate. So, at the early stage, the process of transformation looks like the KJMA one with $n = 4$. It is seen from Eq. (9b) that $\varphi_2(x,\tau) \approx 2\tau^3$ and $Q_3^{(pt)}(\tau) = Q_s(\tau) \approx \exp(-2\alpha_s \tau^3)$ at $\tau >> 1$, i.e. transformation at large times proceeds similarly to the plane transformation with $n = 3$; the multiplier 2 reflects the fact that the plate surface consists of two planes, each of them can equally transform the random point $O'$ in the plate.

The full range of the Avrami exponent variation including the asymptotics $n \to 3$ can be observed only at small values of the parameter $\alpha_s$; small $\alpha_s$ occurs, in particular, at small $R_0$, when the plate indeed looks like a plane. In the case of a large $\alpha_s$, transformation is completed at $\tau \sim 1$ (the new-phase layer formed after the BS reaches the plate centre for this time), hence, we can restrict ourselves by the interval $\tau < 1$. The time dependences of the Avrami exponent, $n^{(pt)}(\tau)$, relatively the limit curve $n_{BS}(\tau)$ are displayed in Fig. 2c; the corresponding curves $n^{(s)}(\tau)$ for a spherical particle [42] are given for comparison (the limit curve $n_{BS}(\tau)$ is the same in both these cases). The plots of $n^{(pt)}(\tau)$ and $n^{(s)}(\tau)$ noticeably diverge at $\alpha_s = 10$ and gradually become closer to each other with increasing $\alpha_s$. The peak origin in Fig. 2b is just the BS phenomenon; the dependence $n^{(pt)}(\tau)$ for a large $\alpha_s$ in this figure is observable experimentally only for $\tau < 1 + \Delta$ (before the maximum).

In view of Eq. (16), the bulk-nucleation KJMA equation corresponding to the considered problem is

$$Q_K(\tau) = \exp\left(-\frac{1}{2}\alpha_s \tau^4\right) \tag{17}$$

The double-logarithmic VF plots together with the corresponding KJMA straight lines are shown in Fig. 2d; the plots are similar to those for spherical particles, in particular, they end with a steep bend at large $\alpha_s$ values.

VF equations for the case of instantaneous (site-saturated) nucleation or/and diffusional growth can be obtained from general Eqs. (6a-d) using the $\delta$ -shaped nucleation rate $I_s(t') = N_s\delta_+(t')$ or/and the growth law $R(t',t) = \sqrt{C(t-t')}$ [42]. VF equations for the case of bulk nucleation in the plate are given by the same Eqs. (6a-d) with replacements $I_s(t') \to I_b(t')$ and $S_i(r;t',t) \to V_i^{(in)}(r;t',t)$, the volume of the CR segment lying inside the plate.

## 3. Ensemble of width-distributed plates

It is assumed here that the parallel plates are *isolated* from each other, so that nuclei of one plate cannot intergrow into adjacent plates (this assumption is similar to that of Johnson-Mehl's model for grain-boundary nucleated transformations [2, 3]). To get the VF of such ensemble, we need the $R_0$ - dependent VF of a plate at a fixed time $t$ [42]. It is obtained from Eqs. (9a-c) by substitutions

$$\rho = \frac{1}{\tau} = \frac{R_0}{ut} = \frac{L}{2ut}, \quad y = \frac{r}{ut}, \quad x = \frac{y}{\rho} \tag{18}$$

The characteristic dimensionless parameter here is $\beta_s = (\pi/3) I_s u^2 t^3$; $\alpha_s = \beta_s \rho^3$. The result is as follows:

(1) $\rho \geq 1$:

$$Q_1^{(pt)}(\beta_s, \rho) = \frac{\rho - 1}{\rho} + \frac{1}{\rho}\int_{\rho-1}^{\rho} \mathrm{e}^{-\beta_s \psi_1(y,\rho)}\, dy, \quad \psi_1(y,\rho) = 1 - 3(\rho - y)^2 + 2(\rho - y)^3 \tag{19a}$$

(2) $\frac{1}{2} < \rho < 1$:

$$Q_2^{(pt)}(\beta_s, \rho) = \frac{1}{\rho}\left\{\int_0^{1-\rho} \mathrm{e}^{-\beta_s \psi_2(y,\rho)}\, dy + \int_{1-\rho}^{\rho} \mathrm{e}^{-\beta_s \psi_1(y,\rho)}\, dy\right\},$$

$$\psi_2(y,\rho) = 2\left[1 - 3(\rho^2 + y^2) + 6\rho y^2 + 2\rho^3\right] \tag{19b}$$

(3) $0 < \rho \leq \frac{1}{2}$:

$$Q_3^{(pt)}(\beta_s, \rho) = \frac{1}{\rho}\int_0^{\rho} \mathrm{e}^{-\beta_s \psi_2(y,\rho)}\, dy \tag{19c}$$

The dependence $X^{(pt)}(\beta_s, \rho) = 1 - Q^{(pt)}(\beta_s, \rho)$ is plotted in Fig. 3a. Differently from the case of a spherical particle, where it has a maximum or plateau (which is inherent in a finite domain [42]), here the dependence $X^{(pt)}(\beta_s, \rho)$ is monotonically decreasing and has the plateau $X(\beta_s,\rho)=1$ at sufficiently large $\beta_s$ s. The BS-limit curve here is $X_{BS}^{(pt)}(\rho)=1/\rho$ for $\rho\geq 1$ and $X_{BS}^{(pt)}(\rho)=1$ for $\rho<1$ (cf. Eq. (13)); the VF curves approach it with increasing $\beta_s$ s.

Let $f_L(L)$ be the probability density function for the plate width. The Erlang distribution

$$f_k^{(L)}(L)=\frac{\lambda(\lambda L)^{k-1}}{(k-1)!}\mathrm{e}^{-\lambda L} \tag{20a}$$

of the $k$ -th order and its particular case of $k=1$ - the exponential distribution

$$f_1^{(L)}(L)=\lambda\,\mathrm{e}^{-\lambda L} \tag{20b}$$

inherent in the Poisson process considered later - are employed hereafter. As is known, the sum of $k$ random quantities distributed by the exponential law, Eq. (20b), is distributed by the $k$ -th order Erlang law, Eq. (20a).

The mean for $f_k^{(L)}(L)$ is $\overline{L}_k=k\lambda^{-1}$, so that $\overline{L}_1=\lambda^{-1}$ for $f_1^{(L)}(L)$. Utilizing $\overline{L}_1$ as a characteristic length, we shall use further the dimensionless width $l=L/\overline{L}_1=L\lambda$ and the dimensionless distributions

$$f_k(l)=\frac{l^{k-1}}{(k-1)!}\mathrm{e}^{-l} \tag{21a}$$

$$f_1(l)=\mathrm{e}^{-l} \tag{21b}$$

The probability for the point $O'$ to fall at random into a plate of the width $[L,L+dL]$ equals to the relative measure of these plates, according to the geometrical definition of probability:

$$p_k(l)dl=\frac{Lf_k^{(L)}(L)dL}{\int_0^\infty Lf_k^{(L)}(L)dL}=\frac{Lf_k^{(L)}(L)dL}{\overline{L}_k}=\frac{lf_k(l)dl}{k}=\frac{l^k\,\mathrm{e}^{-l}\,dl}{k!} \tag{22a}$$

$$p_1(l)dl=l\,\mathrm{e}^{-l}\,dl \tag{22b}$$

These basic probabilities will be reused below.

Multiplying the above VF $Q^{(pt)}(\beta_s,\rho)$ by $p_k(l)dl$ and integrating over the whole range of $l$, we get the desired untransformed VF of the ensemble. Preliminarily, $\beta_s$ and $\rho$ are represented as follows [42]:

$$b\equiv\frac{\pi}{3}\frac{I_s}{u},\quad \overline{\alpha}_s=b\overline{R}_0^3,\quad \overline{R}_0=\overline{L}_1/2,\quad \overline{\tau}=\frac{ut}{\overline{R}_0}=\frac{2ut}{\overline{L}_1}=2ut\lambda,\quad \beta_s=\overline{\alpha}_s\overline{\tau}^3,\quad \rho=\frac{R_0}{ut}=\frac{L}{2ut}=\frac{l}{\overline{\tau}} \tag{23}$$

Finally,

$$Q_k^{(en)}(\overline{\tau})=\int_0^\infty Q^{(pt)}\left(\overline{\alpha}_s\overline{\tau}^3,\frac{l}{\overline{\tau}}\right)p_k(l)dl,\quad X_k^{(en)}(\overline{\tau})=1-Q_k^{(en)}(\overline{\tau}) \tag{24}$$

The BS-limit VF is

$$Q_{BS}^{(pt)}(\rho) = \frac{\rho - 1}{\rho} = 1 - \frac{\bar{\tau}}{l} = Q_{BS}^{(pt)}(l, \bar{\tau}),\ l > \bar{\tau} \tag{25a}$$

which is the untransformed part $1 - 2ut/L$ of the plate of width $L$; $Q_{BS}^{(pt)}(l, \bar{\tau}) = 0$ for $l < \bar{\tau}$. The probability for the point $O'$ to fall into the untransformed part of any plate, i.e. into the untransformed part of the ensemble is

$$Q_k^{(BS)}(\bar{\tau}) = \int_{\bar{\tau}}^{\infty} Q_{BS}^{(pt)}(l, \bar{\tau}) p_k(l) dl \tag{25b}$$

In the two main variants considered here, $k = 1$ and $k = 2$, this gives

$$Q_1^{(BS)}(\bar{\tau}) = \int_{\bar{\tau}}^{\infty} \left(1 - \frac{\bar{\tau}}{l}\right) l\, \mathrm{e}^{-l} dl = \mathrm{e}^{-\bar{\tau}} \tag{26a}$$

$$Q_2^{(BS)}(\bar{\tau}) = \frac{1}{2} \int_{\bar{\tau}}^{\infty} \left(1 - \frac{\bar{\tau}}{l}\right) l^2 \mathrm{e}^{-l} dl = \left(1 + \frac{\bar{\tau}}{2}\right) \mathrm{e}^{-\bar{\tau}} \tag{26b}$$

Obviously, Eqs. (9a-c) can be also interpreted as the untransformed VF of the ensemble of identical plates representing a regular 1D structure. Fig. 3b shows how the width distribution *slows down* the TK as compared to the TK in the corresponding regular structure; the ensemble of identical plates with $L = \bar{L}_1$ and hence $\alpha_s = \bar{\alpha}_s$, $\tau = \bar{\tau}$ is employed for comparison. It is seen that the late-stage TK of the ensemble of exponentially distributed plates at $\bar{\alpha}_s = 10^3$ is even slower than the TK of the ensemble of identical plates at $\alpha_s = 0.1$; the exponential tail of the former corresponds to $n = 1$, whereas $n \to 3$ for the latter, as shown above. Kinetics slows down to a greater extent with the increasing order $k$ of Erlang's distribution. Each kind of curves in Fig. 3b approach its BS-limit line corresponding to Eqs. (13), (26a) and (26b), at large $\alpha_s$, $\bar{\alpha}_s$ values. Obviously, the kinetics slows down due to the contribution of the plates with $L > \bar{L}_1$.

## 4. Infinite number of parallel planes

### 4.1. Regular arrangement

Here the periodic 1D structure of parallel planes with the same spacing $L$ between them is considered, Fig. 4. The sequential addition of new pairs of planes in Fig. 1 is accompanied by increase in the number of time intervals in the VF equation, i.e. in the number of $Q_i(\tau)$, as well as by complication of the latter. Therefore, we should employ some approximation to get a VF equation for the system of an infinite number of planes. Here the same approximation is used, as in Ref. [42] for the model of grain-boundary nucleated transformations – it can be called the *mean-field* approximation (MFA). Marking out the pair containing the random point $O'$ (Fig. 4), the influence of other (external) planes on transformation of the space bounded by this pair is represented as some effective bulk nucleation from the outside. Specifically, the computational procedure is as follows.

We specify the CR at time $t'$ for the point $O'$ taken at random at time $t$ - the sphere of radius $R(t',t)$, as before - and denote by $V_1(r;t',t)$ and $V_2(r;t',t)$ the volumes of spherical segments cut off by the right and left planes, respectively (Fig. 4);

$$V_1(r;t',t) = \frac{\pi}{3}\left[2R^3(t',t) - 3(R_0 - r)R^2(t',t) + (R_0 - r)^3\right],$$

$$V_2(r;t',t) = \frac{\pi}{3}\left[2R^3(t',t) - 3(R_0 + r)R^2(t',t) + (R_0 + r)^3\right] \tag{27}$$

The volumes $V_i^{(in)}(r;t',t)$ mentioned at the end of Section 2 relate to $V_i(r;t',t)$ as follows: $V_1^{(in)}(r;t',t) = (4\pi/3)R^3(t',t) - V_1(r;t',t)$, $V_2^{(in)}(r;t',t) = (4\pi/3)R^3(t',t) - V_1(r;t',t) - V_2(r;t',t)$.

The mean area $S_i^{(ex)}(r;t',t)$ of external planes within the segment of volume $V_i(r;t',t)$ is

$$S_i^{(ex)}(r;t',t) = \omega V_i(r;t',t) \tag{28a}$$

where $\omega$ is the mean area of planes in unit volume. We have for nucleation on external planes

$$I_s S_i^{(ex)}(r;t',t) = I_s \omega V_i(r;t',t) = I_b V_i(r;t',t), \quad I_b = \omega I_s \tag{28b}$$

where $I_b$ is the effective bulk nucleation rate. This equation means that nucleation on external planes can be represented as bulk nucleation outside the given pair of planes.

Calculating the probability $Q(r,t)$ for the point $O'$ to be untransformed at time $t$ according to the procedure of Section 2, we get the functions $Y_i^{(ex)}(r,t)$ representing the contribution of external planes, similarly to the functions $Y_i(r,t)$, as follows:

$$Y_1^{(ex)}(r,t) = \begin{cases} 0, & 0 \le r \le r_0 \\ \int_0^{t_m(r,t)} I_b(t')V_1(r;t',t)dt', & r_0 < r \le R_0 \end{cases} \tag{29a}$$

$$Y_2^{(ex)}(r,t) = \begin{cases} \int_0^{t'_m(r,t)} I_b(t')[V_2(r;t',t) + V_1(r;t',t)]dt' + \int_{t'_m(r,t)}^{t_m(r,t)} I_b(t')V_1(r;t',t)dt', & 0 \le r < r'_0 \\ \int_0^{t_m(r,t)} I_b(t')V_1(r;t',t)dt', & r'_0 \le r \le R_0 \end{cases} \tag{29b}$$

$$Y_3^{(ex)}(r,t) = \int_0^{t'_m(r,t)} I_b(t')[V_2(r;t',t) + V_1(r;t',t)]dt' + \int_{t'_m(r,t)}^{t_m(r,t)} I_b(t')V_1(r;t',t)dt', \quad 0 \le r < R_0 \tag{29c}$$

The mentioned probability $Q_i(r,t)$ is determined by the sum

$$Y_i^{(tot)}(r,t) = Y_i(r,t) + Y_i^{(ex)}(r,t) \tag{30}$$

Calculation of $Y_i^{(ex)}(r,t)$ in the isothermal case with the use of the above dimensionless variables yields the characteristic parameter

$$\alpha_b = \frac{\pi}{3}\frac{I_b}{u} R_0^4 = c\alpha_s\,, \quad c \equiv \omega R_0$$

so that one obtains

(1) $\tau < 1$:

$$Q_1^{(pl)}(\tau) = (1-\tau) + \int_{1-\tau}^{1} \mathrm{e}^{-\alpha_s[\varphi_1(x,\tau)+c\varphi_1^{(ex)}(x,\tau)]}\,dx \tag{31a}$$

(2) $1 \le \tau \le 2$:

$$Q_2^{(pl)}(\tau) = \int_{0}^{\tau-1} \mathrm{e}^{-\alpha_s[\varphi_2(x,\tau)+c\varphi_2^{(ex)}(x,\tau)]}\,dx + \int_{\tau-1}^{1} \mathrm{e}^{-\alpha_s[\varphi_1(x,\tau)+c\varphi_1^{(ex)}(x,\tau)]}\,dx \tag{31b}$$

(3) $\tau > 2$:

$$Q_3^{(pl)}(\tau) = \int_{0}^{1} \mathrm{e}^{-\alpha_s[\varphi_2(x,\tau)+c\varphi_2^{(ex)}(x,\tau)]}\,dx \tag{31c}$$

where the functions $\varphi_i(x,\tau)$ are given by Eqs. (9a, b), as before, whereas the functions $\varphi_i^{(ex)}(x,\tau)$ represent the contribution of external planes:

$$\varphi_1^{(ex)}(x,\tau) = \frac{1}{2}\tau^4 - (1-x)\tau^3 + (1-x)^3\tau - \frac{1}{2}(1-x)^4 \tag{32a}$$

$$\varphi_2^{(ex)}(x,\tau) = \tau^4 - 2\tau^3 + 2(1+3x^2)\tau - (1+6x^2+x^4) \tag{32b}$$

For $n$ pairs of planes, $\omega = 2n/(2n-1)L$, which gives $\omega = 1/L$ in the limit $n \to \infty$. Hence, $c = R_0/L = 1/2$.

This is the 1D regular model of grain-boundary nucleated transformations. Fig. 5a displays the kinetic curves $X^{(pl)}(\tau)$ in comparison with the VFs $X^{(pt)}(\tau)$ for the ensemble of identical isolated plates, Eqs. (9a-c), for different $\alpha_s$ s. As it was expected, $X^{(pl)}(\tau) > X^{(pt)}(\tau)$ due to the effect of external planes; this effect is noticeable at small $\alpha_s$ s and diminishes with their increase due to the developing BS phenomenon. The latter counteracts the influence of external planes, so that both the dependences converge; it is seen that they are close already at $\alpha_s = 1$.

### 4.2. Random arrangement: the mean field approximation

Let the spacing $L$ between planes be a random quantity with the probability density $f_k^{(L)}(L)$. This case differs from the above case of the ensemble of width-distributed plates in that the “plates” are not isolated now, i.e. a nucleus born on any plane contributes to the transformation of the whole system (a growing nucleus can cross planes). This is the 1D random model of grain-boundary nucleated transformations. For calculating the VF here, the above time-dependent VF, Eqs. (31a-c), must be transformed to a radius-dependent VF, similarly to Eqs. (19a-c).

Substituting Eqs. (18) in Eqs. (31a-c), we get the characteristic parameter $\beta_b = (\pi/3)I_b u^3 t^4$, so that $\alpha_b = \beta_b \rho^4$ and

$$\beta_b = \beta_s \omega u t = \beta_s \omega \sqrt[3]{\beta_s / b}\,, \quad b \equiv \frac{\pi}{3}\frac{I_s}{u} \tag{33a}$$

Expressions of the form $\beta_s[\psi_i(y,\rho) + \omega\sqrt[3]{\beta_s/b}\,\psi_i^{(ex)}(y,\rho)]$ are obtained as arguments of exponential functions. Eq. (33a) can be further transformed using the characteristic length $\overline{L}_1$ and Eq. (23):

$$\omega\sqrt[3]{\beta_s/b} = \omega\overline{R}_0\overline{\tau} = \overline{c}\,\overline{\tau}\,, \quad \overline{c} = \omega\overline{R}_0 \tag{33b}$$

As a result,

(1) $\rho \ge 1$:

$$Q_1^{(pl)}(\beta_s,\rho,\overline{\tau}) = \frac{\rho-1}{\rho} + \frac{1}{\rho}\int_{\rho-1}^{\rho} \mathrm{e}^{-\beta_s[\psi_1(y,\rho)+\overline{c}\,\overline{\tau}\psi_1^{(ex)}(y,\rho)]}\,dy \tag{34a}$$

(2) $\frac{1}{2} < \rho < 1$:

$$Q_2^{(pl)}(\beta_s,\rho,\overline{\tau}) = \frac{1}{\rho}\left\{\int_0^{1-\rho} \mathrm{e}^{-\beta_s[\psi_2(y,\rho)+\overline{c}\,\overline{\tau}\psi_2^{(ex)}(y,\rho)]}\,dy + \int_{1-\rho}^{\rho} \mathrm{e}^{-\beta_s[\psi_1(y,\rho)+\overline{c}\,\overline{\tau}\psi_1^{(ex)}(y,\rho)]}\,dy\right\} \tag{34b}$$

(3) $0 < \rho \le \frac{1}{2}$:

$$Q_3^{(pl)}(\beta_s,\rho,\overline{\tau}) = \frac{1}{\rho}\int_0^{\rho} \mathrm{e}^{-\beta_s[\psi_2(y,\rho)+\overline{c}\,\overline{\tau}\psi_2^{(ex)}(y,\rho)]}\,dy \tag{34c}$$

where the functions $\psi_i(y,\rho)$ are given by Eqs. (19a, b), whereas the functions $\psi_i^{(ex)}(y,\rho)$ are as follows:

$$\psi_1^{(ex)}(y,\rho) = \frac{1}{2} - (\rho - y) + (\rho - y)^3 - \frac{1}{2}(\rho - y)^4 \tag{35a}$$

$$\psi_2^{(ex)}(y,\rho) = 1 + 2\rho(\rho^2 + 3y^2 - 1) - (\rho^4 + 6y^2\rho^2 + y^4) \tag{35b}$$

Similarly to Eq. (24), the desired VF equation for the considered system in MFA is

$$Q_k^{(mf)}(\overline{\tau}) = \int_0^\infty Q^{(pl)}\left(\overline{\alpha}_s\overline{\tau}^3, \frac{l}{\overline{\tau}}, \overline{\tau}\right) p_k(l)dl \qquad (36)$$

It will be shown later that

$$\omega = \lambda = \overline{L}_1^{-1} \qquad (37)$$

for the exponential distribution $f_1(l)$; hence, $\overline{c} = \omega\overline{R}_0 = \overline{R}_0 / \overline{L}_1 = 1/2$ in this case. However, $\overline{c}$ can be different for another application of Eqs. (34a-c) considered later.

It should be noted that the influence of external planes discussed above is absent in the BS limit, so that the same results take place for the ensemble of width-distributed plates and for the ensemble of random planes; in particular, Eqs. (26a, b) hold for random planes as well.

Fig. 5b shows the kinetic curves $X_1^{(mf)}(\overline{\tau})$ in comparison with the VFs $X_1^{(en)}(\overline{\tau})$ for the ensemble of width-distributed isolated plates, Eq. (24), for different $\alpha_s$s. Both qualitatively and quantitatively, the picture is the same, as in Fig. 5a, and the reasons are the same as well. Only the kinetics is slowed in whole as compared to Fig. 5a due to the exponential distribution, as shown in Fig. 3b.

### 4.3. Random arrangement: exact solution

The system of random parallel planes with the Poisson distribution in space allows an exact solution for the VF which can be obtained by the method of Ref. [7]. This fact is valuable to estimate the MFA accuracy.

Preliminarily, we find the probability $q_{pl}(r,t)$ for the point $O'$ to be untransformed at time $t$ from the plane at a distance $r$ from it. Evidently, the condition $r < R_m(t)$ must be satisfied; otherwise, $q_{pl}(r,t) = 1$. As follows from the foregoing, the result is

$$q_{pl}(r,t) = \exp\left[-\int_0^{t_m(r,t)} I_s(t')S(r;t',t)dt'\right], \quad S(r;t',t) = \pi[R^2(t',t) - r^2] \qquad (38a)$$

where $t_m(r,t)$ is determined by the equality $R(t_m,t) = r$. In the isothermal case, this gives

$$q_{pl}(x,t) = \mathrm{e}^{-\beta_s(t)(1-3x^2+2x^3)}, \quad x = r/R_m(t) = r/ut, \quad \beta_s(t) = (\pi/3)I_s u^2 t^3 \qquad (38b)$$

As before, we take at random a point $O'$ in the considered system, Fig. (6), and find the probability $Q_{(1)}(t)$ that it is untransformed at time $t$. The CR for the point $O'$ at time $t' = 0$ is the sphere of radius $R_m(t)$. Let it be intersected by $n$ planes. The diametrical line intersects these planes at points $C_1$, $C_2$, ..., $C_n$ representing the Poisson point process (P-process).

Denoting by $r_k = |O'C_k|$ the distance from the point $O'$ to the $k$ th plane, $k = 1, 2, \ldots, n$, the probability $q_n(\{r_k\},t)$ for the point $O'$ to be untransformed at time $t$ at the given $n$ and realization of the set $\{r_k\} = \{r_1, r_2, ..., r_n\}$ is found as a product of the probabilities $q_{pl}(r_k,t)$, Eq. (38a):

$$q_n(\{r_k\},t) = \prod_{k=1}^{n} q_{pl}(r_k,t) = \prod_{k=1}^{n} \exp\left[-\int_0^{t_m(t,r_k)} I_s(t')S(r_k;t',t)dt'\right] \qquad (39a)$$

This equation reflects the possibility for a growing nucleus to cross planes, which is the property of Cahn's model [38]. Due to this property, the transformations of the point $O'$ from different planes are independent events, which results in the product ().

The desired function $Q(t)$ is obtained by averaging $q_n(\{r_k\},t)$ over all $r_k$ - values and $n$:

$$Q_{(1)}(t) = \sum_{n=0}^{\infty} P_{(1)}(n) \int_0^{R_m(t)} \ldots \int_0^{R_m(t)} q_n(\{r_k\},t)\phi(r_1, r_2, ..., r_n)dr_1 dr_2 \ldots dr_n, \quad P_{(1)}(n) = \frac{(\overline{n})^n}{n!}\mathrm{e}^{-\overline{n}} \qquad (39b)$$

where $\phi(\{r_k\})$ is the distribution density of the set $\{r_k\}$ and $P_{(1)}(n)$ is the Poisson distribution. The positions $r_k$ of the points $C_k$ are *independent* random quantities *uniformly* distributed on the segment $[AB]$, Fig. 6, at the condition that their number on this segment is $n$ [50], which is an important property of the P-process. According to the independency, $\phi(r_1, r_2, ..., r_n) = \tilde{\phi}(r_1)\tilde{\phi}(r_2)\ldots\tilde{\phi}(r_n)$. According to the uniformity, the probability for the point $C_k$ to be in the interval $dr_k$ equals to $dr_k / 2R_m$, where $2R_m$ is the length of the segment $[AB]$. With respect to the point $O'$, the point $C_k$ can be at the right or at the left at the same distance $r_k$ (two equal possibilities), hence, this probability is doubled and $\tilde{\phi}(r_k)dr_k = dr_k / R_m$, $\tilde{\phi}(r_k) = 1/R_m$. As a result, the $n$-fold integral in Eq. (39b) equals to $w(t)^n$, where

$$w(t) = \frac{1}{R_m(t)} \int_0^{R_m(t)} q_{pl}(r,t)dr = \int_0^1 q_{pl}(x,t)dx \qquad (40)$$

Averaging with the Poisson distribution yields

$$Q_{(1)}(t) = \sum_{n=0}^{\infty} \frac{\overline{n}^n}{n!}\mathrm{e}^{-\overline{n}}\, w(t)^n = \mathrm{e}^{-\overline{n}(1-w(t))} \qquad (41)$$

The mean number $\overline{n}$ of points on the segment $[AB]$ of length $2R_m(t)$ in the P-process with the intensity $\lambda$ is

$$\overline{n} = 2\lambda R_m(t) \qquad (42)$$

The mean area of the plane part located inside the CR is

$$\overline{S} = \pi \int_0^{R_m(t)} (R_m^2 - r^2)\frac{dr}{R_m} = \frac{2}{3}\pi R_m^2 \tag{43a}$$

Hence, the mean area of planes in unit volume is

$$\omega = \frac{\overline{S}\overline{n}}{(4\pi/3)R_m^3} = \frac{\overline{n}}{2R_m} \tag{43b}$$

from where

$$\overline{n} = 2\omega R_m(t) \tag{44}$$

Comparing to Eq. (42), we get the equality $\omega = \lambda$ used above, Eq. (37).

Finally, isothermal Eq. (41) has the following explicit form:

$$Q_{(1)}(t) = \exp\left\{-2\omega ut\int_0^1 \left[1 - \mathrm{e}^{-\beta_s(t)(1-3x^2+2x^3)}\right]dx\right\} \tag{45a}$$

which is the Cahn equation [38, 41]. The result is clear, as the basic probability $q_{pl}(r,t)$ depends only on the distance $r$ to a plane, but not on its orientation. The given case can be considered as the 1D Cahn model. Cahn's method being an application of the classical JMA approach with the extended volume concept is valid here just to the *Poisson distribution of planes*, similarly to usual (point) nucleation, where both approaches – mathematically rigorous Kolmogorov's and inventive JMA's – give the same result. As it will be shown in the next Section, Cahn's method fails in the case of a non-Poissonian distribution of planes.

Denoting, as before, $\overline{\tau} = 2\omega ut = 2ut\lambda$ and $\beta_s = \overline{\alpha}_s\overline{\tau}^3$, we get

$$Q_{(1)}(\overline{\tau}) = \exp\left\{-\overline{\tau}\int_0^1 \left[1 - \mathrm{e}^{-\overline{\alpha}_s\overline{\tau}^3(1-3x^2+2x^3)}\right]dx\right\}, \quad X_{(1)}(\overline{\tau}) = 1 - Q_{(1)}(\overline{\tau}) \tag{45b}$$

From the above equations, the extended volume $X_e^{(1)}(\overline{\tau})$ of Cahn's theory [38] can be represented as

$$X_e^{(1)}(\overline{\tau}) = \overline{\tau}(1 - w(\overline{\tau})) \quad w(\overline{\tau}) = \int_0^1 \mathrm{e}^{-\overline{\alpha}_s\overline{\tau}^3(1-3x^2+2x^3)}\,dx \tag{46}$$

In the BS limit, $\overline{\alpha}_s \to \infty$, one obtains $w(\overline{\tau}) \to 0$ and

$$X_{e,BS}^{(1)}(\overline{\tau}) = \overline{\tau}\,, \quad Q_{(1)}^{(BS)}(\overline{\tau}) = \mathrm{e}^{-\overline{\tau}}\,, \quad X_{(1)}^{(BS)}(\overline{\tau}) = 1 - \mathrm{e}^{-\overline{\tau}} \tag{47}$$

in accordance with Eq. (26a). The exponential in these equations takes into account the overlap of growing plates; the quantity $\overline{\tau}$ can be also interpreted as the volume of growing and overlapping plates in unit volume of the system, by definition. This overlap can be neglected at $\overline{\tau} \to 0$; then $X_{(1)}^{(BS)}(\overline{\tau}) = X_{e,BS}^{(1)}(\overline{\tau}) = \overline{\tau}$.

It should be also noted that, according to Eq. (42), $\bar{n}=\bar{\tau}$ in the isothermal process, so that $P_{(1)}(n)=P_{(1)}(n,\bar{\tau})$.

### 4.4. Random arrangement: non-Poissonian distribution

A non-Poissonian distribution of planes is of great interest; apparently it was not considered in literature as yet in connection with TK. Intervals between the successive events (points) in the P-process are independent and have the same exponential distribution $f_1^{(L)}(L)$, Eq. (20b). An evident and important generalization is obtained under the assumption that these intervals are independent and have the same distribution $f^{(L)}(L)$ different from the exponential one. The resulting series of point events on the time axis is called the *renewal process*; these processes are studied in the *renewal theory* [51, 52] which is a section of the theory of reliability of technical devices. One of the instances is the $k$ th-order Erlang point process ($E_k$-process), $k>1$; it is obtained from the P-process as result of its thinning out, when each $k$ th event of the P-process is retained and all other events are removed.

As a simplest example of a non-Poissonian distribution of planes, the $E_2$-process is considered here; it is obtained as a result of thinning out the P-process, when each second plane is removed. The intervals between the remaining planes (similar to the *inter-arrival times* in the renewal theory) are distributed according to the second-order Erlang law $f_2^{(L)}(L)$, Eq. (20a) with $k$ =2. As follows from the previous Section, all we need to derive the VF $Q_{(2)}(t)$ in this case is an expression for $P_{(2)}(n)$, the probability of $n$ points on the given segment $[AB]$, and the distribution $\phi(\{r_k\})$ of these points.

The probability $P_n(x)$ is calculated in Appendix. The segment $[0,x)$ used in calculations is the segment $[AB]$ here, i. e. $x=2R_m(t)=2ut$, $\lambda x=2ut\lambda=\bar{\tau}$. The mean $\bar{n}(x)$ in the considered process, as is shown in Appendix, equals to $\lambda x/2=\lambda R_m(t)$; comparing this equation to Eq. (44) which is obviously valid in the given case as well, we get

$$\omega=\frac{\lambda}{2} \tag{48}$$

for the $E_2$-process, which is an expected result, since the number of planes is twice less than in the parent P-process.

In view of the above relations, Eq.(A14) acquires the following form

$$P_{(2)}(n)=P_{(2)}(n,\bar{\tau})=\mathrm{e}^{-\bar{\tau}}\left\{\frac{1}{2}\frac{\bar{\tau}^{2n+1}}{(2n+1)!}+\frac{\bar{\tau}^{2n}}{(2n)!}+\frac{1}{2}\frac{\bar{\tau}^{2n-1}}{(2n-1)!}\right\},\quad n=1,\,2,\,3,\,\ldots \tag{49a}$$

$$P_{(2)}(0,\bar{\tau}) = \mathrm{e}^{-\bar{\tau}}\left(1+\frac{1}{2}\bar{\tau}\right) \tag{49b}$$

Each of the points of the parent P-process on the segment $[AB]$ is uniformly distributed on it; in addition, the positions of all these points are independent. Obviously, the $n$ points remaining after the thinning procedure retain both of these properties; hence, the $n$-fold integral in Eq. (39b) equals to $w(t)^n$, as before, where $w(t)$ and $w(\bar{\tau})$ are given by Eqs. (40) and (46). In this way,

$$Q_{(2)}(\bar{\tau}) = \sum_{n=0}^{\infty} P_{(2)}(n,\bar{\tau})w^n = P_{(2)}(0,\bar{\tau}) + \mathrm{e}^{-\bar{\tau}}\left\{\frac{1}{2\sqrt{w}}\sum_{n=1}^{\infty}\frac{(\sqrt{w}\bar{\tau})^{2n+1}}{(2n+1)!} + \sum_{n=1}^{\infty}\frac{(\sqrt{w}\bar{\tau})^{2n}}{(2n)!} + \frac{\sqrt{w}}{2}\sum_{n=1}^{\infty}\frac{(\sqrt{w}\bar{\tau})^{2n-1}}{(2n-1)!}\right\} \tag{50a}$$

These are series for hyperbolic sine and cosine:

$$Q_{(2)}(\bar{\tau}) = \mathrm{e}^{-\bar{\tau}}\left(1+\frac{1}{2}\bar{\tau}\right) + \mathrm{e}^{-\bar{\tau}}\left\{\frac{1}{2\sqrt{w}}\left[\sinh(\sqrt{w}\bar{\tau}) - \sqrt{w}\bar{\tau}\right] + \left[\cosh(\sqrt{w}\bar{\tau}) - 1\right] + \frac{\sqrt{w}}{2}\sinh(\sqrt{w}\bar{\tau})\right\} \tag{50b}$$

Finally,

$$Q_{(2)}(\bar{\tau}) = \mathrm{e}^{-\bar{\tau}}\left\{\frac{1}{2}\left[\sqrt{w(\bar{\tau})} + \frac{1}{\sqrt{w(\bar{\tau})}}\right]\sinh\left(\sqrt{w(\bar{\tau})}\,\bar{\tau}\right) + \cosh\left(\sqrt{w(\bar{\tau})}\,\bar{\tau}\right)\right\} \tag{51}$$

In the BS limit, we have $w \to 0$, $\sinh(\sqrt{w}\bar{\tau})/\sqrt{w} \to \bar{\tau}$, and $\cosh(\sqrt{w}\bar{\tau}) \to 1$, so that the preexponential factor becomes equal to $(1+\bar{\tau}/2)$ and Eq. (51) goes to Eq. (26b), as it must.

It is seen that Eqs. (26a) and (26b) coincide with the probabilities $P_{(1)}(0,\bar{\tau})$ and $P_{(2)}(0,\bar{\tau})$ of the absence of points on the segment $[AB]$ in the P- and E$_2$-process, respectively. Equivalently, these are probabilities that no plane intersects the CR. Indeed, the plane located at the distance $R_m(t) = ut$ from the point $O'$ (at either the point $A$ or $B$) transforms it just at time $t$ in the BS limit, whereas a plane located at $r_k < ut$ transforms the point $O'$ before time $t$. Hence, the CR has not to be intersected by planes to keep the point $O'$ untransformed at time $t$ in the BS limit. This result can be directly applied to the site-saturated nucleation on a straight line with the density $\Lambda$ of nucleation sites. With the notation $\tilde{\tau} = 2\Lambda ut$, the KJMA equation is $X_K(\tilde{\tau}) = 1 - \exp(-\tilde{\tau})$ in this case. If the nucleation sites represent the E$_2$ process, then the VF equation is $X_E(\tilde{\tau}) = 1 - (1+\tilde{\tau}/2)\exp(-\tilde{\tau})$.

Denoting by $\chi(\bar{\tau})$ the preexponential expression in Eq. (51), we have

$$X_{(2)}(\bar{\tau}) = 1 - Q_{(2)}(\bar{\tau}) = 1 - \exp\left[-X_e^{(2)}(\bar{\tau})\right],\quad X_e^{(2)}(\bar{\tau}) \equiv \bar{\tau} - \ln\chi(\bar{\tau}) \tag{52a}$$

$$\frac{dX_{(2)}(\bar{\tau})}{d\bar{\tau}} = \left(1 - X_{(2)}(\bar{\tau})\right)\frac{dX_e^{(2)}(\bar{\tau})}{d\bar{\tau}} \tag{52b}$$

These are familiar equations of the JMA extended-volume approach [2-5, 38]; $X_e^{(2)}(\bar{\tau})$ is the "extended volume" for the E2-process. In the BS limit, $X_{e,BS}^{(2)}(\bar{\tau}) = \bar{\tau} - \ln(1+\bar{\tau}/2)$, from where $X_{e,BS}^{(2)}(\bar{\tau}) = X_{(2)}^{(BS)}(\bar{\tau}) = \bar{\tau}/2$ at $\bar{\tau} \to 0$, which is twice less than $X_{(1)}^{(BS)}(\bar{\tau}) = X_{e,BS}^{(1)}(\bar{\tau}) = \bar{\tau}$ in the P-process due to the twice less number of planes here.

The extended volume $X_e^{(1)}(\bar{\tau})$ expression, Eq. (46), was derived by Cahn [38] from clear geometrical constructions and it has a proper interpretation; as was mentioned above, $X_{e,BS}^{(1)}(\bar{\tau}) = \bar{\tau}$ is the volume of overlapping plates in unit volume of the system, in accordance with the extended volume definition. Similarly, the extended volume in the JMA theory is the volume of overlapping spherical nuclei. It is seen that the "extended volume" $X_{e,BS}^{(2)}(\bar{\tau})$ of the E2-process involves a logarithmic summand and therefore does not have the same interpretation as $X_{e,BS}^{(1)}(\bar{\tau})$; it is only clear that $X_{e,BS}^{(2)}(\bar{\tau}) < X_{e,BS}^{(1)}(\bar{\tau})$ due to the more rare arrangement of planes. The $X_e^{(2)}(\bar{\tau})$ expression is much more complex than $X_e^{(1)}(\bar{\tau})$ one and has no clear geometrical interpretation, differently from the latter; therefore, it cannot be derived by the method of work [38]. So, Cahn's method is successful only in the P-process and not applicable to non-Poissonian ones. On the contrary, the present method as a more rigorous and general one allows us to get exact solutions for these processes as well.

Fig. 7a shows the kinetic curves $X_{(1)}(\bar{\tau})$ and $X_{(2)}(\bar{\tau})$ at small and large $\alpha_s$ s. TK is slower for the E2-process, as it must. At large $\alpha_s$ s, each kind of curves approach its BS-limit line $X_{(k)}^{BS}(\bar{\tau})$, Eqs. (26a, b).

Figs. 7b and 7c give a comparison of the exact, $X_{(k)}(\bar{\tau})$, and MFA, $X_k^{(mf)}(\bar{\tau})$, curves for the P- and E2-process, respectively. As it can be seen, the MFA accuracy is surprisingly good for the P-process and somewhat worse for the E2-one; MFA equations slightly overestimate the exact kinetics. One can say that the MFA accuracy for the E2-process "lags" from that for the P-process by one order of magnitude in $\alpha_s$, i.e. the accuracy in the former case at $\alpha_s = 1$ is about the same as the accuracy in the latter case at $\alpha_s = 0.1$. The accuracy improves with increasing $\alpha_s$ values obviously due to the developing BS phenomenon; when the effect of external planes diminishes, the MFA and exact solutions give close results. The error given by the MFA in each process, $\Delta X_{(k)}(\bar{\tau}) = X_k^{(mf)}(\bar{\tau}) - X_{(k)}(\bar{\tau})$, in the region of small $\alpha_s$ s is shown in Fig. 7d. The error behavior is not monotonic with $\alpha_s$ change; $\Delta X_{(k)}(\bar{\tau})$ has a maximum at $\alpha_s \sim 10^{-3}$. The maximal error in VF cubes $\Delta X_{(1)}^{(c)}(\bar{\tau}) = (Q_{(1)}(\bar{\tau}))^3 - (Q_1^{(mf)}(\bar{\tau}))^3$ is about 0.01; this fact will be used later.

## 5. Structures of random parallelepipeds: paradox of packing

### 5.1. Basic structure

Three sets of random parallel planes orthogonal to each other form the grain structure consisting of random parallelepipeds (hereafter, pp-s for brevity), Fig. 8a; let $L_x, L_y$, and $L_z$ denote the pp edges. The randomness of pp-s should be clarified in more detail. If we take *at random* one pp, its edges ($L_x, L_y$, $L_z$) are *independent random* quantities. However, all the pp-s forming one plate between the neighboring planes, say, $XY$, have the same $L_z$-value. Similarly, the pp-s forming rectangular cylinders have the same values of pairs, say, ($L_x, L_y$). In other words, there is a *strong correlation* in the arrangement of pp-s here; this fact is crucial for further consideration.

To get the VF of untransformed material in this structure, we take at random the point $O'$ and calculate the probability $Q_{(a)}(t)$ that it is untransformed at time $t$. Evidently, $Q_{(a)}(t)$ is found as the product of three probabilities to be untransformed in each set of parallel planes, in view of the independence of these events. If the planes in each of three sets are distributed according to the P-process with the same parameter $\lambda$, then $Q_{(a)}^{(1)}(t) = Q_{(1)}(t)^3$, where $Q_{(1)}(t)$ is given by Eq. (45a), and

$$Q_{(a)}^{(1)}(\bar{\tau}) = Q_{(1)}(\bar{\tau})^3 = \exp\left\{-3\bar{\tau}\int_0^1 \left[1 - \mathrm{e}^{-\bar{\alpha}_s \bar{\tau}^3 (1-3x^2+2x^3)}\right] dx\right\},\quad X_{(a)}^{(1)}(\bar{\tau}) = 1 - Q_{(a)}^{(1)}(\bar{\tau}) \tag{53}$$

i. e. the area $\omega_{pp}$ of boundaries in unit volume here is three times greater: $\omega_{pp} = 3\omega$. Similarly, $Q_{(a)}^{(2)}(\bar{\tau}) = Q_{(2)}(\bar{\tau})^3$ for the $E_2$-process.

Having this 3D structural model, we can compare it to 3D Cahn's model (with arbitrary orientation of planes). The measure element of planes in their parametric space ($r$, $\theta$, $\varphi$) is [53] $dM_C = \sin\theta\, dr d\theta d\varphi \equiv dr d\Omega$, where $d\Omega = \sin\theta\, d\theta d\varphi$ is the solid angle element. The probability for a plane to be at a distance $r_k$ from the point $O'$ in 3D can be obtained from this measure. Indeed, the measure of all planes located in the interval $[r_k, r_k + dr_k]$ equals to $dM_{r_k} = 4\pi dr_k$, whereas the total measure of planes intersecting the CR is $M_C = 4\pi R_m$. The mentioned probability is found as a relative measure: $dM_{r_k} / M_C = dr_k / R_m$.

Having the P-process of planes in the space ($r$, $\theta$, $\varphi$) with the intensity $\lambda_C$, the probability to find a plane in the measure element $\Delta M_C$ is $\lambda_C \Delta M_C + o(\Delta M_C)$. The mean number of planes intersecting the CR is $\bar{n} = \lambda_C M_C = 4\pi R_m \lambda_C$. On the other hand, Eq. (44) with $\omega_C$ instead of $\omega$ holds in the general Cahn model as well [7], $\bar{n} = 2\omega_C R_m$. From comparison,

$$\omega_C = 2\pi\lambda_C \tag{54a}$$

If we put $\omega_C$ the same as in the considered model of random pp-s, $\omega_C = \omega_{pp} = 3\lambda$, then one obtains

$$\lambda_C = \frac{3}{2\pi}\lambda \tag{54b}$$

i. e. for the same area of boundaries in unite volume as in the model of pp-s, the "density" of planes in the general Cahn model must be approximately twice less than the "density" of planes in the 1D Cahn model; this result is due to the additional measure $\Omega = 4\pi$ of planes connected with orientation.

The volume distribution of pp-s in the given model is of interest; it will be essentially used later. Having a pp ($L_x, L_y$, $L_z$) randomly chosen in the given structure, the task is to find the distribution of the product $L_x L_y L_z = V$ for known distributions $f^{(L)}(L_i)$ of the quantities $L_i$, $i = x, y, z$. At first, we find the distribution function $G(V)$ as the probability for the point ($L_x, L_y$, $L_z$) to be under the surface $L_x L_y L_z = V$, by definition:

$$G(V) = P\{L_x L_y L_z < V\} = \int_0^\infty dL_x \int_0^\infty dL_y \int_0^{V/L_x L_y} dL_z f^{(L)}(L_x) f^{(L)}(L_y) f^{(L)}(L_z) \tag{55a}$$

Further, the desired probability density function $g^{(V)}(V)$ is obtained by differentiating:

$$g^{(V)}(V) = \frac{dG(V)}{dV} = \int_0^\infty dL_x \int_0^\infty dL_y \frac{1}{L_x L_y} f^{(L)}(L_x) f^{(L)}(L_y) f^{(L)}\left(\frac{V}{L_x L_y}\right) \tag{55b}$$

We can take the most general probability densities $f^{(L)}(L_i)$ as the $k_i$ th order Erlang laws with different $\lambda_i$ s:

$$f^{(L)}(L_i) = f_{k_i}^{(L)}(L_i) = \frac{\lambda_i (\lambda_i L_i)^{k_i - 1}}{(k_i - 1)!} \mathrm{e}^{-\lambda_i L_i}, \quad i = x, y, z \tag{56}$$

The above VF $Q_{(a)}(t)$ can be easily obtained for all combinations with $k_i = 1, 2$ and different $\lambda_i$ s as a product of the corresponding 1D VFs. Combining different $k_i$ s with different $\lambda_i$ s in Eq. (56), we can model numerous pp structures with the exact probability density $g^{(V)}(V)$ for them. Any real grain distribution can be also fitted by the function $g^{(V)}(V)$ with properly chosen parameters $k_i$ and $\lambda_i$. Structures with strongly different $\lambda_i$s or/and $k_i$s will contain pp-s with strongly different $L_i$, e.g., the plate-like pp-s ($L_x, L_y >> L_z$) or wire-like ones ($L_x, L_y << L_z$).

Here $\lambda_i = \lambda$ and $k_i = k$ are used for simplicity; Eq. (55b) in this case is as follows:

$$g^{(V)}(V) = \frac{\lambda^3}{((k-1)!)^3} \int_0^\infty dL_x \int_0^\infty dL_y \frac{(\lambda^3 V)^{k-1}}{L_x L_y} \mathrm{e}^{-\lambda\left[L_x + L_y + \frac{V}{L_x L_y}\right]} \tag{57a}$$

Going to the dimensionless variables $l_i = \lambda L_i$, $\upsilon = \lambda^3 V$, $g(\upsilon) = \lambda^{-3} g^{(V)}(V)$, one obtains finally

$$g_k(\upsilon) = \frac{1}{((k-1)!)^3} \int_0^\infty dl_x \int_0^\infty dl_y \frac{\upsilon^{k-1}}{l_x l_y} \mathrm{e}^{-\left[l_x + l_y + \frac{\upsilon}{l_x l_y}\right]} \tag{57b}$$

Defining the length $l_c$ by relation $\upsilon = l_c^3$, the distribution over $l_c$ is obtained as

$$g_k^{(l_c)}(l_c) = 3 l_c^2 g_k(l_c^3) \tag{57c}$$

## 5.2. Derivative structures

### 5.2.1. Preliminary remarks

In what follows, an alternative way for deriving the VF equation is employed. The probability $Q_{(a)}(t)$ that the random point $O'$ at time $t$ falls into the untransformed part of the system is obtained as the averaging product of two probabilities: (i) the probability $dQ_{pp}(L_x, L_y, L_z)$ for the point $O'$ to fall into a pp with the edges $[L_i, L_i + dL_i]$, $i = x, y, z$, and (ii) the probability $q_{pp}(L_x, L_y, L_z; t)$ to fall into the untransformed part of this pp. Let the parallel planes forming the basic structure of Fig. 8a (hereafter called the (a)-structure) be distributed according to the P-process with the same parameter $\lambda$; the considered further derivative structures retain this property (the superscript (1) is omitted).

In this way, one obtains for the (a)-structure in the dimensionless variables $l_i = \lambda L_i$, $\bar{\tau} = 2ut\lambda$

$$dQ_{pp}^{(a)}(l_x, l_y, l_z) = p_1(l_x) p_1(l_y) p_1(l_z) dl_x dl_y dl_z \tag{58a}$$

$$Q_{(a)}(\bar{\tau}) = \int_0^\infty dl_x \int_0^\infty dl_y \int_0^\infty dl_z q_{pp}(l_x, l_y, l_z; \bar{\tau}) l_x l_y l_z \, \mathrm{e}^{-(l_x + l_y + l_z)} \tag{58b}$$

This integral is the probability for the point $O'$ to fall into the untransformed part of any pp, i. e. into the untransformed part of the whole system. In view of the independence condition,

$$q_{pp}(l_x, l_y, l_z; \bar{\tau}) = q_{pp}(l_x, \bar{\tau}) q_{pp}(l_y, \bar{\tau}) q_{pp}(l_z, \bar{\tau}) \tag{59}$$

for the (a)-structure, where $q_{pp}(l_i,\overline{\tau})$ is the conditional probability for the point $O'$ to be untransformed in the set of parallel planes at the condition that it is located between two planes separated by the distance $l_i$. Here equations for $q_{pp}(l_i,\overline{\tau})$ are the same, hence

$$Q_{(a)}(\overline{\tau}) = \left( \int_0^{\infty} q_{pp}(l,\overline{\tau}) l\, \mathrm{e}^{-l}\, dl \right)^3 \tag{60}$$

For the structures considered below, either Eq. (59) does not hold or equations for $q_{pp}(l_i,\overline{\tau})$ are not the same and calculating some of them is a complicated procedure resulting in a cumbersome expression. However, such exact or MFA equations for $q_{pp}(l_i,\overline{\tau})$ are unnecessary for understanding the main idea of the present approach, so that the BS limit is further employed for $q_{pp}(l_x,l_y,l_z;\overline{\tau})$. In this limit,

$$q_{pp}(L_x,L_y,L_z;t) = \begin{cases} \dfrac{(L_x-2ut)(L_y-2ut)(L_z-2ut)}{L_xL_yL_z}, & \eta(L_x-2ut)\eta(L_y-2ut)\eta(L_z-2ut)=1 \\ 0, & \eta(L_x-2ut)\eta(L_y-2ut)\eta(L_z-2ut)=0 \end{cases} \tag{61a}$$

where $\eta(x)$ is the Heaviside step function; the first condition means that $L_x, L_y, L_z > 2ut$, whereas $L_i < 2ut$ at least for one $L_i$ in the second condition. In the dimensionless variables,

$$q_{pp}(l_x,l_y,l_z;\overline{\tau}) = \begin{cases} \left(1-\dfrac{\overline{\tau}}{l_x}\right)\left(1-\dfrac{\overline{\tau}}{l_y}\right)\left(1-\dfrac{\overline{\tau}}{l_z}\right), & \eta(l_x-\overline{\tau})\eta(l_y-\overline{\tau})\eta(l_z-\overline{\tau})=1 \\ 0, & \eta(l_x-\overline{\tau})\eta(l_y-\overline{\tau})\eta(l_z-\overline{\tau})=0 \end{cases} \tag{61b}$$

So, in the BS limit both the separation given by Eq. (59) always holds and equations for $q_{pp}(l_i,\overline{\tau})$ are the same; they have a simplest form, $q_{pp}(l_i,\overline{\tau}) = 1-\overline{\tau}/l_i$.

Eq. (60) gives in this limit

$$Q_{(a)}(\overline{\tau}) = \left( \int_{\overline{\tau}}^{\infty} \left(1-\frac{\overline{\tau}}{l}\right) l\, \mathrm{e}^{-l}\, dl \right)^3 = \mathrm{e}^{-3\overline{\tau}} \tag{62}$$

in accordance with Eq. (26a); this result is also obtained from Eq. (53) at $\alpha_s \to \infty$. In this Section, the superscript (BS) is omitted for brevity.

The derivative structures obtained from the (a)-one are as follows.

### 5.2.2. Structures with one and two degrees of freedom

(b)-Structure is obtained from the (a)-one by shifting the rectangular cylinders ($L_x$, $L_z$) relatively each other along the $Y$-axis at random distances, Fig. 8b. As a result, only two sets of parallel planes, $XY$ and $YZ$, remain and there is no more the set of parallel planes $XZ$. Now the above probability $dQ_{pp}(L_x, L_y, L_z)$ for the point $O'$ to fall into the pp ($L_x$, $L_y$, $L_z$) is calculated differently from case (a). The probabilities of $L_x$ and $L_z$ are calculated, as before: $p_1(L_x)p_1(L_z)dL_x dL_z$ - this is the probability for the point $O'$ to fall into the rectangular cylinder ($L_x$, $L_z$). The probability of the remaining quantity $L_y$ does not equal to $p_1(L_y)dL_y$, since there is no parallel planes $XZ$ here; this is simply the probability for the random quantity $L_y$ to be in the interval $[L_y, L_y + dL_y]$ which is determined as $f_1(L_y)dL_y$ by definition. The difference from case (a) is as follows. Consider the subset of the same cylinders ($L_x$, $L_z$). The point $O'$ being in any of them has the *same* $L_y$-value in case (a) and *different* $L_y$-values in case (b); the probabilities of these values in each case are given above. This means that transition from (a)- to (b)-structure weakens the correlation in the arrangement of pp-s mentioned above by one step (with respect to $L_y$ only) and therefore the (b)-structure is a structure with one "degree of freedom". Finally,

$$dQ_{pp}^{(b)}(l_x, l_y, l_z) = p_1(l_x)p_1(l_z)dl_x dl_z f_1(l_y)dl_y \tag{63a}$$

and

$$Q_{(b)}(\bar{\tau}) = \left(\int_{\bar{\tau}}^{\infty}\left(1-\frac{\bar{\tau}}{l}\right)l\,\mathrm{e}^{-l}\,dl\right)^2 \int_{\bar{\tau}}^{\infty}\left(1-\frac{\bar{\tau}}{l_y}\right)\mathrm{e}^{-l_y}\,dl_y = \mathrm{e}^{-2\bar{\tau}}[\mathrm{e}^{-\bar{\tau}} + \bar{\tau}\,\mathrm{Ei}(-\bar{\tau})],\ \ \mathrm{Ei}(-\bar{\tau}) = \int_{-\infty}^{-\bar{\tau}}\frac{\mathrm{e}^x}{x}dx \tag{63b}$$

where $\mathrm{Ei}(x)$ is the integral exponent.

Fig. 8c shows the formation of (c)-structure from the base (a)-one: the plates between the neighboring planes $XY$ are shifted relatively each other along both the $X$ and $Y$ axes at random distances. As a result, only one set of parallel planes, $XY$, remains. In this way, the probability of $L_z$ is calculated here as $p_1(L_z)dL_z$ and the probabilities of $L_x$ and $L_y$ as $f_1(L_x)f_1(L_y)dL_x dL_y$. Consider the subset of plates with the same width $L_z$. The point $O'$ being in any of them has the *same* $L_x$, $L_y$-values in case (a); indeed, to get the desired pp ($L_x$, $L_y$, $L_z$) ), we take the cylinder ($L_x$, $L_y$) and cut off its parts of width $L_z$ by planes $XY$. At the same time, the point $O'$ has *different* pairs ($L_x$, $L_y$) in case (c). In this way, transition from (a)- to (c)-structure weakens the correlation in the arrangement of pp-s by two steps (with respect to both $L_x$ and $L_y$) and therefore the (c)-structure is a structure with two "degrees of freedom". With respect to the (b)-structure, the correlation is weakened by one step more.

Finally,

$$dQ_{pp}^{(c)}(l_x, l_y, l_z) = p_1(l_z)dl_z f_1(l_x)f_1(l_y)dl_x dl_y \tag{64a}$$

$$Q_{(c)}(\bar{\tau}) = \int_{\bar{\tau}}^{\infty}\left(1-\frac{\bar{\tau}}{l_z}\right)l_z\,\mathrm{e}^{-l_z}\,dl_z\left(\int_{\bar{\tau}}^{\infty}\left(1-\frac{\bar{\tau}}{l}\right)\mathrm{e}^{-l}\,dl\right)^2 = \mathrm{e}^{-\bar{\tau}}[\mathrm{e}^{-\bar{\tau}}+\bar{\tau}\,\mathrm{Ei}(-\bar{\tau})]^2 \tag{64b}$$

### 5.2.3. Completely uncorrelated structure

As follows from the above consideration, the remaining step in the weakening of correlation is to destroy the set of parallel planes of the (c)-structure. In this way, we decompose the base (a)-structure into single pp-s and then repack them randomly, Fig. 8d; this is the (d)-structure. As there are no parallel planes, a length $l$ (the spacing between planes) is no longer a basic measure for calculating the probability for the random point $O'$ to fall into a certain region; the pp volume $\upsilon$ is a basic measure here. The probability for the point $O'$ to fall into a pp of volume $\upsilon$ is the relative measure of these pp-s, by definition:

$$P_1(\upsilon)d\upsilon = \frac{\upsilon g_1(\upsilon)d\upsilon}{\int_0^{\infty}\upsilon g_1(\upsilon)d\upsilon} \tag{65}$$

where $g_1(\upsilon)$ is given by Eq. (57b) with $k=1$.

Having a given volume $\upsilon$, the pp can have two arbitrary edges $l_x$ and $l_y$ with the probabilities $f_1(l_x)dl_x$ and $f_1(l_y)dl_y$; the third edge is $l_z = \upsilon / l_x l_y$. As before, the untransformed part $q_{pp}(l_x, l_y, l_z; \bar{\tau})$ of the pp at time $\bar{\tau}$ depends on these lengths:

$$q_{pp}(l_x,l_y,\upsilon;\bar{\tau}) = \begin{cases} \dfrac{1}{\upsilon}(l_x-\bar{\tau})(l_y-\bar{\tau})\left(\dfrac{\upsilon}{l_xl_y}-\bar{\tau}\right), & \eta(l_x-\bar{\tau})\eta(l_y-\bar{\tau})\eta\left(\dfrac{\upsilon}{l_xl_y}-\bar{\tau}\right)=1 \\ 0, & \eta(l_x-\bar{\tau})\eta(l_y-\bar{\tau})\eta\left(\dfrac{\upsilon}{l_xl_y}-\bar{\tau}\right)=0 \end{cases}, \tag{66}$$

In this way,

$$dQ_{pp}^{(d)}(l_x,l_y,\upsilon) = P_1(\upsilon)d\upsilon\, f_1(l_x)f_1(l_y)dl_xdl_y \tag{67a}$$

$$Q_{(d)}(\bar{\tau}) = \int_{\bar{\tau}}^{\infty}dl_x\int_{\bar{\tau}}^{\infty}dl_yf_1(l_x)f_1(l_y)\int_{\bar{\tau}l_xl_y}^{\infty}q_{pp}(l_x,l_y,\upsilon;\bar{\tau})P_1(\upsilon)d\upsilon$$

$$= \int_{\bar{\tau}}^{\infty}dl_x\int_{\bar{\tau}}^{\infty}dl_y(l_x-\bar{\tau})(l_y-\bar{\tau})\mathrm{e}^{-(l_x+l_y)}\left\{\frac{1}{l_xl_y}\int_{\bar{\tau}l_xl_y}^{\infty}P_1(\upsilon)d\upsilon-\bar{\tau}\int_{\bar{\tau}l_xl_y}^{\infty}\frac{P_1(\upsilon)d\upsilon}{\upsilon}\right\} \tag{67b}$$

Evidently, the present approach with the probability $P_1(\upsilon)$ cannot be applied to the (a)-structure for calculating $dQ_{pp}^{(a)}$, where the probability for the point $O'$ to fall into any pp of volume $\upsilon$ is *different* – pp-s of the same volume $\upsilon$ have different sets $(l_x, l_y, l_z)$ and the probability $dQ_{pp}^{(a)}(l_x,l_y,l_z)$ is determined just by these sets, rather than by $\upsilon$, Eq. (58a). The same holds for the (b)- and (c)-structures. The converse statement is also true: the approach of case (a) is not applicable to case (d), since there are no parallel planes here. Among the considered structures, evidently, the (d)-structure corresponds to a real grain structure. Just the difference in equations for the probabilities $dQ_{pp}^{(d)}$ and $dQ_{pp}^{(a)}$ underlies the difference in the real TK and that of Cahn's model, as is discussed below.

The kinetic curves $X_{(\alpha)}(\overline{\tau}) = 1 - Q_{(\alpha)}(\overline{\tau})$, $\alpha =$ a, b, c, d, are shown in Fig. 9a. It is seen that the sequential weakening of correlation from (a) to (d) accelerates the TK; it is the fastest in the completely uncorrelated (d)-structure and the slowest in the (a)-structure which is a particular case of Cahn's model. Interestingly that curve (c) is very close to curve (d); the maximal difference between them is about $8\times10^{-3}$, which means that the correlation already in the (c)-structure can be regarded weak.

In this way, the structures consisting of the same pp-s, but differently packed, give different rates of transformation; this phenomenon is called here the paradox of packing. It explains why the general Cahn model underestimates the real TK [46]. Although the planes in this model are at random angles, there is a strong correlation in the arrangement of polyhedra, similarly to the (a)-structure – each array of polyhedra is tied to a certain plane; each polyhedron belongs to several such arrays, as each pp in the (a)-structure belongs to three arrays of pp-s. As shown above, the Cahn equation with $\omega_C = 3\lambda$ is the same as the (a)-structure VF equation, i.e. the particular and general Cahn models with the same $\omega_C$ value give the same TK and therefore are indistinguishable from the physical point of view. The exponential asymptotics of Cahn's equation corresponding to the 1D growth ($n = 1$) originates just from random planes and therefore is closely related to the mentioned correlation. The real transformation curves (at a finite $\alpha_s$) approach those of Fig. 9a with increasing $\alpha_s$, as in previous figures.

It is seen from the foregoing that a mathematical reason for the difference in kinetic curves (a)-(d) is different probabilistic measures $dQ_{pp}^{(\alpha)}$, similarly to the classical Bertrand paradox [53]. From the physical point of view, different patterns of space filling occur in these cases. The space is filled by three sets of growing plates in case (a), by two sets of plates and single transforming pp-s in case (b), by one set of plates and single pp-s in case (c), and only by single transforming pp-s in case (d). To understand how single transforming pp-s accelerate the TK, the (C)- and (B)-structures corresponding to the (c)- and (b)-ones can be considered. The (C)-structure is simply the system of parallel planes with the exponential distribution of spacing between them, i. e. when the plates in the (c)-structure are not divided into pp-s; as shown above, the BS-limit kinetic curve here is $X_{(C)}(\bar{\tau}) = 1 - \exp(-\bar{\tau})$. If rectangular cylinders in the (b)-structure are not divided into pp-s, we get the (B)-structure; obviously, $X_{(B)}(\bar{\tau}) = 1 - \exp(-2\bar{\tau})$. Comparing $X_{(b)}(\bar{\tau})$ to $X_{(B)}(\bar{\tau})$ and $X_{(c)}(\bar{\tau})$ to $X_{(C)}(\bar{\tau})$ in Fig. 9a, we see how the presence of pp-s accelerates the TK; the difference between curves (C) and (c) is especially sharp. As curve (c) is close to (d)-one, this comparison sheds light on the result obtained for the (d)-structure.

### 5.2.4. Intermediate structures

In addition to the main (a)-(d)-structures described above, different intermediate structures can be considered as well. One of them, the (c')-structure, is shown in Fig. 8c'; the difference from the (c)-structure is in that the blocks of two adjacent plates are shifted relatively each other, rather than single plates. Of course, this classification is conditional: if extreme cases (a) and (d) are considered as main, then the (b)- and (c)-structures are also intermediate.

Let $l = \lambda L$ is the block width, Fig. 8c'; it is distributed according to the second-order Erlang law. The probability for the point $O'$ to fall into the pp $(l_x, l_y, l)$ is given by equation similar to Eq. (64a), but with $p_2(l)dl$, Eq. (22a) with $k = 2$, instead of $p_1(l_z)dl_z$:

$$dQ_{pp}^{(c')}(l_x, l_y, l) = p_2(l)dl f_1(l_x) f_1(l_y) dl_x dl_y, \quad p_2(l) = \frac{l^2\,\mathrm{e}^{-l}}{2} \tag{68a}$$

Hence, Eq. (64b) is replaced by the following one:

$$Q_{(c')}(\bar{\tau}) = c'(\bar{\tau})[\mathrm{e}^{-\bar{\tau}} + \bar{\tau}\,\mathrm{Ei}(-\bar{\tau})]^2 \tag{68b}$$

where $c'(\bar{\tau})$ is the probability for the point $O'$ to be in the untransformed part of the length $l$; in the case when the length $l$ is not separated into two parts, $c'(\bar{\tau})$ is given by Eq. (26b). In the considered case, calculations are more complicated.

The untransformed part of the length $l$ is calculated differently for different $l_z$-values of one of the pp-s constituting the block; the possible three variants are shown in Fig. 10. The quantity $l_z$ is limited now by the condition $l_z \le l$, so that its distribution $f_1(l_z)$ must be normalized in accordance with this condition:

$$\tilde{f}_1(l_z, l) = z(l)\mathrm{e}^{-l_z}, \quad z(l) = \left(\int_0^l \mathrm{e}^{-l_z}\, dl_z\right)^{-1} = \left(1 - \mathrm{e}^{-l}\right)^{-1} \tag{69}$$

The mentioned three variants are as follows.

(i) $0 < L_z < 2ut$ $\;(0 < l_z < \bar{\tau})$

The untransformed part here is

$$\frac{L - L_z - 2ut}{L} = 1 - \frac{l_z + \bar{\tau}}{l} \tag{70a}$$

In this way, the probability of variant (i) is

$$c'_{(i)}(\bar{\tau}) = \int_0^{\bar{\tau}} dl_z \, \mathrm{e}^{-l_z} \int_{l_z+\bar{\tau}}^{\infty} dl \left(1 - \frac{l_z + \bar{\tau}}{l}\right) p_2(l) z(l) \tag{70b}$$

(ii) $2ut < L_z < L - 2ut$ ($\bar{\tau} < l_z < l - \bar{\tau}$)

In this case, the untransformed region consists of two parts:

$$\frac{(L - L_z - 2ut) + (L_z - 2ut)}{L} = 1 - \frac{2\bar{\tau}}{l} \tag{71a}$$

Changing the order of integration,

$$\int_{2\bar{\tau}}^{\infty} dl \int_{\bar{\tau}}^{l-\bar{\tau}} dl_z = \int_{\bar{\tau}}^{\infty} dl_z \int_{l_z+\bar{\tau}}^{\infty} dl$$

one obtains

$$c'_{(ii)}(\bar{\tau}) = \int_{\bar{\tau}}^{\infty} dl_z \, \mathrm{e}^{-l_z} \int_{l_z+\bar{\tau}}^{\infty} dl \left(1 - \frac{2\bar{\tau}}{l}\right) p_2(l) z(l) \tag{71b}$$

(iii) $L - 2ut < L_z < L$ ($l - \bar{\tau} < l_z < l$)

Going to the new variable $l'_z = l_z - (l - \bar{\tau})$, $0 < l'_z < \bar{\tau}$, we get the untransformed part as follows:

$$\frac{L_z - 2ut}{L} = \frac{l_z - \bar{\tau}}{l} = 1 - \frac{2\bar{\tau} - l'_z}{l} \tag{72a}$$

The probability of this variant is

$$c'_{(iii)}(\bar{\tau}) = \int_0^{\bar{\tau}} dl'_z \, \mathrm{e}^{-l'_z} \int_{2\bar{\tau}-l'_z}^{\infty} dl \left(1 - \frac{2\bar{\tau} - l'_z}{l}\right) \mathrm{e}^{-(l-\bar{\tau})} p_2(l) z(l) \tag{72b}$$

Finally, the desired function $c'(\bar{\tau})$ is the sum of these three,

$$c'(\bar{\tau}) = c'_{(i)}(\bar{\tau}) + c'_{(ii)}(\bar{\tau}) + c'_{(iii)}(\bar{\tau}) \tag{73}$$

It turns out that $c'_{(ii)}(\bar{\tau})$ is a monotonically decreasing function (from 1 to 0), whereas $c'_{(i)}(\bar{\tau})$ and $c'_{(iii)}(\bar{\tau})$ are bell-shaped with maximum values 0.3 and 0.04, respectively. In this way, the first two summands give the main contribution to the sum; the smallness of $c'_{(iii)}(\bar{\tau})$ is evidently due to large $l_z$- values in this case and hence the exponentially small their probabilities.

Similarly, the (b')-structure can be formed, if the $2\times2$ blocks of rectangular cylinders are shifted relatively each other in Fig. 8b, rather than single cylinders; let $l^{(x)} = \lambda L^{(x)}$ and $l^{(z)} = \lambda L^{(z)}$ be the block sizes distributed according to the second-order Erlang law. The probability for the point $O'$ to fall into the pp $(l^{(x)}, l_y, l^{(z)})$ is

$$dQ_{pp}^{(b')}(l^{(x)},l_y,l^{(z)}) = p_2(l^{(x)})p_2(l^{(z)})dl^{(x)}dl^{(z)}f_1(l_y)dl_y \tag{74a}$$

and

$$Q_{(b')}(\bar{\tau}) = (c'(\bar{\tau}))^2[\mathrm{e}^{-\bar{\tau}} + \bar{\tau}\,\mathrm{Ei}(-\bar{\tau})] \tag{74b}$$

Numerous intermediate structures can be constructed in the similar way by increasing the block sizes (as well as considering $k \times m$ blocks in case (b'), $k$ - and $m$ -blocks in case (c'), etc.), however, the complexity of analytical calculations also increases. All these structures differ from each other by the degree of correlation in the arrangement of pp-s. E.g., the degree of correlation in the structure of $k \times k$ blocks which generalizes the (b')-one increases with increasing $k$ and approaches that of the base (a)-structure. In this way, the degree of correlation in the (c')- and (b')-structures is slightly greater, than in the corresponding (c)- and (b)-structures, hence, the TK should be slightly slower. At first glance, it seems that the kinetic curves $X_{(c')}(\bar{\tau})$ and $X_{(b')}(\bar{\tau})$ coincide with $X_{(c)}(\bar{\tau})$ and $X_{(b)}(\bar{\tau})$, respectively in Fig. 9a. However, there are still the small differences $\Delta X_{(c-c')} = X_{(c)}(\bar{\tau}) - X_{(c')}(\bar{\tau})$ and $\Delta X_{(b-b')} = X_{(b)}(\bar{\tau}) - X_{(b')}(\bar{\tau})$ between them shown in Fig. 9b. The maximum of $\Delta X_{(c-c')}$ is about $10^{-3}$, whereas the $\Delta X_{(b-b')}$ maximum is four times greater. The smallness of $\Delta X_{(c-c')}$ was expected due to the small correlation in the (c)-structure itself. As is mentioned above, these differences will increase with increasing the block sizes.

## 6. Cubic structures

### 6.1. Regular cubic structure and its derivatives

A *regular* cubic structure is formed by three sets of parallel planes considered in Section 4.1 and similar to the base (a)-structure of Fig. 8a. Obviously, the MFA untransformed VF in this structure is

$$Q_i^{(c)}(\tau) = [Q_i^{(pl)}(\tau)]^3 \tag{75}$$

as a consequence of the independency of events for the point $O'$ to be untransformed from each set; $Q_i^{(pl)}(\tau)$ is given by Eqs. (31a-c). This equation can be also derived directly extending the procedure of Section 2 successively to 2D and 3D spaces.

From this base (a)-structure, we can form the (b)-(d)-structures in the same manner, as in the previous Section. To understand the TK in these structures, the interpretation of Fig. 4 should be somewhat changed. Let this figure depicts the transformation of a single plate under the influence of some external surfaces. The parallel planes as such surfaces are employed in Fig. 4, however, the result, Eqs. (31a-c), depends only on the area $\omega$ of these surfaces in unit volume; hence, we can use *arbitrary surfaces* instead of planes in the MFA. Three mutually orthogonal plates, each with its environment, form a cube with some resulting environment; the VF of this cube in the MFA is given by Eq. (75). Let $\omega^{(c)}$ be the area of boundaries in unit volume for the (a)-(d) structures; evidently, $\omega^{(c)} = 3\omega = 3/L = 3/2R_0$ for the base (a)-structure. Being an average quantity, by definition,

$$\omega^{(c)} = \lim_{V\to\infty} \frac{S(V)}{V} \tag{76}$$

it has the same value for the (a)-(c) structures, despite the fact that cubes have different surrounding surfaces in these structures. Due to the independence of this environment stated above, the MFA TK in these structures is *the same* and given by Eq. (75). In particular,

$$Q_{BS}^{(c)}(\tau) = \begin{cases} (1-\tau)^3, & \tau \le 1 \\ 0, & \tau > 1 \end{cases} \tag{77}$$

in the BS limit, when cubes are “isolated” from each other.

The (d)-structure consisting of randomly packed cubes may contain a void between them depending on the packing algorithm. The condition of minimum void can be imposed to get the random dense packing of cubes. The presence of a void affects the $\omega^{(c)}$ value for this structure and therefore the TK; with the use of $\omega^{(c)}$ as an input parameter, Eq. (75) with $\omega = \omega^{(c)}/3$ in equations for $Q_i^{(pl)}(\tau)$ describes the TK in this structure as well. So, the results given below for a regular cubic structure take place for any system of identical cubes with the same $\omega^{(c)}$ value.

The effect of a void on the $\omega^{(c)}$ value can be exemplified by the following two cases. (i) Consider the spheres of radius $R_0 = L/2$ inscribed in the cubes of the base (a)-structure; they form the lattice of spheres dual to the given cubic lattice. It was shown [42] that $c^{(s)} = \omega^{(s)}R_0 = \pi/2 = 1.57$ for this spherical lattice, whereas $c^{(c)} = \omega^{(c)}R_0 = 1.5$ for the given (a)-structure. We see that $\omega^{(s)} > \omega^{(c)}$ despite the fact that the surface area of a sphere is almost twice less than the surface area of the corresponding cube and there is a void between the spheres. This result takes place due to the fact that one face belongs to two adjacent cubes, so that $S(V) = N(V)(6L^2)/2$, where $N(V)$ is the number of cubes in volume $V = N(V)L^3$; i.e. the total area is divided by two giving $\omega^{(c)} = 3/L$.

(ii) Repack the base (a)-structure into the new structure, where the cubes touch each other at the edges, rather than at the faces. Thus we have two sets of cubes nested within each other, one of them is the set of cubic voids. Differently from the previous case, the total surface area is not divided by two here, however, the total volume of the system is twice larger; as a result, we have the same $\omega^{(c)} = 3/L$ value. These two instances show that the presence of a void in the system does not necessarily lead to a decrease in the $\omega^{(c)}$ value.

Fig. 10a shows kinetic curves for regular and the corresponding random planes with the exponential distribution (the 1D Cahn model), both were calculated in the MFA; the correspondence is provided by the equality $\lambda^{-1} = \overline{L}_1 = L$, from where $\overline{\tau} = \tau$ and $\overline{\alpha}_s = \alpha_s$ . Cahn's curves largely underestimate those for regular planes; the divergence between them increases with increase in $\alpha_s$. At $\alpha_s = 10^4$, each curve practically coincides with its BS-limit one, $X_{BS}^{(pl)}(\tau) = \tau$ for $\tau \le 1$, $X_{BS}^{(pl)}(\tau) = 1$ for $\tau > 1$ and $X_{1,BS}^{(mf)}(\tau) = 1 - \exp(-\tau)$. These equations explain the mentioned divergence: all Cahn's curves end with the exponential tail, which is not the case of regular-planes curves.

Fig. 10b displays kinetic curves for the regular cubic structure (or its derivative structures, in view of the above arguments), $c^{(c)} = 1.5$, together with the curves for the dual structure of spheres ($c^{(s)} = 1.57$) [42] and the corresponding Cahn equation which is Eq. (53) due to the equalities $L = \lambda^{-1}$, $\omega^{(c)} = 3/L = 3\lambda$; just this $\omega^{(c)}$ value is substituted into Cahn's equation giving $2\omega^{(c)}ut = 2(3\lambda)ut = 3\overline{\tau}$. The curves for cubes and spheres are close already at $\alpha_s \sim 0.1$; the difference between them vanishes at large $\alpha_s$s. This difference becomes noticeable only in the region of small $\alpha_s$, where its behavior is not monotonic, as in Fig. 7d; the maximal divergence is 0.08 at $\alpha_s \sim 10^{-3}$. The limit $\alpha_s \to 0$ corresponds, in particular, to $R_0 \to 0$, which means that the system becomes homogeneous and both the types of curves should follow the KJMA kinetics in this limit [28, 29, 42]. It turns out that the condition of homogeneity is more inherent in spherical geometry; the kinetic curve for spheres practically coincide with the KJMA one $X_K(\tau) = 1 - \exp(-c^{(s)}\alpha_s\tau^4)$ already at $\alpha_s \sim 10^{-3}$, whereas the corresponding curve for cubes still slightly overestimates these two.

As the kinetic curves for cubes were calculated in the MFA, it would be reasonable to use the MFA Cahn curves for comparison. However, as is seen from Fig. 7d (bottom solid lines), the difference between the MFA an exact Cahn curves is less than 0.01 even at small $\alpha_s$s, so that exact Eq. (53) is employed. Cahn's model noticeably underestimates the TK in the cubic structure. As is mentioned above, this occurs due to the quite different long-time asymptotics in both the models; the exponential asymptotics with $n \to 1$ takes place in Cahn's model. In the cubic structure, $n \approx 4$ at small $\alpha_s$s (as in the case of spheres [42]) and $n$ increases sharply (Fig. 2) for Eq. (77) at large $\alpha_s$s.

### 6.2. Random cubic structures

Developing further the present approach, we can obviously apply it to the (d)-structure with size-distributed cubes; Eqs. (34a-c) are used for this purpose. Having

$$Q_{i,c}(\beta_s,\rho,\bar{\tau}) = [Q_i^{(pl)}(\beta_s,\rho,\bar{\tau})]^3 \qquad (78)$$

for a single cube of size $\rho$ in the system, one obtains for the whole system [42]

$$Q^{(c)}(\bar{\tau}) = \frac{\int_0^{\infty} Q_c(\bar{\alpha}_s\bar{\tau}^3,\frac{l}{\bar{\tau}},\bar{\tau})l^3 f(l)dl}{\int_0^{\infty} l^3 f(l)dl} \qquad (79)$$

where $f(l)$ is the size distribution of cubes.

The area of boundaries in unit volume is

$$\omega^{(c)} = \frac{\frac{6}{2}\int_0^{\infty} L^2 f(L)dL + \frac{1}{2}S_v}{\int_0^{\infty} L^3 f(L)dL + V_v} \qquad (80a)$$

where $V_v$ and $S_v$ are the volume of voids and their total surface area, respectively; these voids are inevitable in a random structure. As is seen from this equation and the above examples, the volume and surface effects of voids on the $\omega^{(c)}$ value counteract to each other, so that the following estimate for $\omega^{(c)}$ is employed here for calculations:

$$\omega^{(c)} = 3\frac{\int_0^{\infty} L^2 f(L)dL}{\int_0^{\infty} L^3 f(L)dL} \qquad (80b)$$

Considering $\omega^{(c)}$ as an input parameter, the $\omega = \omega^{(c)}/3$ value is used in the equations for $Q_i^{(pl)}(\beta_s,\rho,\bar{\tau})$. For the regular structure, $f(L) = \delta(L)$, Eq. (80b) gives $\omega^{(c)} = 3/L$.

The quantity $\omega^{(c)}$ is not needed in the BS limit, where Eq. (79) is as follows:

$$Q_{i,BS}^{(c)}(\bar{\tau}) = \frac{\int_{\bar{\tau}}^{\infty}\left(1-\frac{\bar{\tau}}{l}\right)^3 l^3 f(l)dl}{\int_0^{\infty} l^3 f(l)dl} \qquad (81)$$

Here, as well as in the above VF equations for parallel planes, only the characteristic length parameter $\lambda^{-1}$ is used to determine the dimensionless quantities.

The first distribution of interest is that corresponding to the random pp-s of Fig. 8a; the correspondence of a pp to the cube of the same volume is given by Eq. (57c). The distributions $g_k(v)$ and $g_k^{(l)}(l)$ are shown in Fig. 11 for some $k$ values; the function $g_1(v)$ is monotonically decreasing, whereas others have a maximum. In this way,

$$\omega_k^{(c)} = 3\lambda \frac{\int_0^\infty l^2 g_k^{(l)}(l)dl}{\int_0^\infty l^3 g_k^{(l)}(l)dl} \equiv 3\lambda\gamma_k\,,\quad \bar{c}_k^{(c)} = \omega_k^{(c)}\bar{R}_0 = \frac{\omega_k^{(c)}}{2\lambda} = \frac{3}{2}\gamma_k \tag{82}$$

and $c = \bar{c}_k^{(c)}/3 = 0.5\gamma_k$ is substituted into Eqs. (34a-c).

To compare the obtained kinetic curves with the corresponding Cahn kinetics, we substitute the above $\omega^{(c)}$ value into the Cahn VF equation. Having $2\omega_k^{(c)}ut = (3\lambda\gamma_k)(2ut) = 3\gamma_k\bar{\tau}$, one obtains

$$Q_C(\bar{\tau}) = \exp\left\{-3\gamma_k\bar{\tau}\int_0^1 \left[1 - \mathrm{e}^{-\bar{\alpha}_s\bar{\tau}^3(1-3x^2+2x^3)}\right]dx\right\} \tag{83}$$

One obtains $\gamma_1 = 0.736$, $\bar{c}_1^{(c)} = 1.104$ and $\gamma_2 = 0.426$, $\bar{c}_2^{(c)} = 0.639$ for $k = 1$ and 2, respectively.

The normal distribution

$$f_n(l) = \frac{1}{\sigma\sqrt{2\pi}}\mathrm{e}^{-\frac{(l-1)^2}{2\sigma^2}} \tag{84}$$

is useful to study the change in the TK depending on the distribution width $\sigma$. One obtains $\gamma_n = 0.958$, $\bar{c}_n^{(c)} = 1.437$ and $\gamma_n = 0.858$, $\bar{c}_n^{(c)} = 1.287$ for $\sigma = 0.15$ and 0.3, respectively; Eq. (83) holds with the replacement $\gamma_k \to \gamma_n$.

Finally, the uniform distribution $f_u(L) = 1/2\bar{L}$ on the segment $[0, 2\bar{L}] = [0, 2\lambda^{-1}]$ with $\bar{L} = \lambda^{-1}$ is of interest as the extreme case of a wide distribution; $f_u(l) = 1/2$. As is mentioned above, the grain size distribution resulting from continuous nucleation and growth is wide and close to the uniform one. Eq. (79) has the following form in this case:

$$Q^{(c)}(\bar{\tau}) = \frac{1}{4}\int_0^2 Q_c(\bar{\alpha}_s\bar{\tau}^3, \frac{l}{\bar{\tau}}, \bar{\tau})l^3 dl \tag{85}$$

Applying Eq. (82) with $f_u(l)$ instead of $g_k^{(l)}(l)$ and with integration up to 2, one obtains $\gamma_u = 2/3$, $\omega_u^{(c)} = 2\lambda$, $\bar{c}_u^{(c)} = 1$, $c = 1/3$, and $2\omega_u^{(c)}ut = 2(2\lambda ut) = 2\bar{\tau}$ for the corresponding Cahn equation.

Figs. 12a-c show the kinetic curves for these four distributions in comparison with other characteristic curves for $\bar{\alpha}_s = 0.1$ and $10^3$. The first common feature of these figures is that the TK in the system of size-distributed cubes is slower than in the corresponding regular cubic structure ( $L=\lambda^{-1}, \tau=\bar{\tau}$ , and $\alpha_s=\bar{\alpha}_s$ ); this is the same phenomenon as in the above case of the ensemble of width-distributed plates, Fig. 3b, and random planes, Fig. 10a. In the case of normal distribution, Fig. 12c, the kinetic curve approaches that for the regular structure when the distribution width $\sigma$ decreases, as it must.

The second common feature is that the Cahn model equation underestimates the obtained kinetics, which agrees with previous studies [42, 46]. As in Fig. 10b, here the exact Cahn equation instead of the MFA one is employed. However, the accuracy correction with the use of Fig. 7d does not change the relative positions of the mentioned curves at $\bar{\alpha}_s=0.1$; this issue is absent at large $\bar{\alpha}_s$ s. It can be seen that the degree of this underestimation is different for different distributions; it is minimal for the function $g_1^{(l)}(l)$ and maximal for the normal distribution. On the other hand, Eq. (53) for the (a)-structure of pp-s corresponding to the given cubic structure overestimates the obtained TK, despite the fact that both structures have the same volume distribution of structural elements. The difference between the (a)-structure curve and Cahn's one is due to the fact that $\gamma_1<1$, i.e. different $\omega$-values; the given-model curve is between these two.

Fig. 12b shows the same curves as Fig. 12a, but for the function $g_2^{(l)}(l)$ corresponding to the second-order Erlang law. Accordingly, kinetics is slower here, than in the previous case, and the curves move further away from the regular-structure graph. This also applies to the curve for the (d)-structure of pp-s (Fig. 8d) which is calculated here for Erlang's distribution as well as the (a)-structure curve. Finally, kinetic curves for the uniform distribution are shown in Fig. 12c. In the cases of normal and uniform distributions, Cahn's equation noticeably underestimates the TK.

Fig. 13 shows the double-logarithmic VF plots (for $\bar{\alpha}_s=10^4$) usually employed by experimentalists for the analysis of TK, and the corresponding temporal behavior of the Avrami exponent. While the positions of all curves in Fig. 13a (except for the BS-limit line) depend on the $\omega^{(c)}$ value (they go down with decrease in $\omega^{(c)}$), their slope does not change; just the latter is important here. Therefore, only the Cahn curve for the $g_2^{(l)}(l)$ distribution as the lowest one and only the KJMA line for the regular ( $\delta$-distributed) cubic structure as the highest one are shown. At the early stage, each curve coincides with the corresponding KJMA line. Further, they diverge; simultaneously, the curves for cubic structures deviate from the Cahn-curve direction. In addition, the curves for $\delta$-shaped and uniform distributions end in a pronounced steep bend, which leads to a sharp increase in the Avrami exponent in Fig. 13b. In the case of small $\bar{\alpha}_s$ s, the Avrami exponent value is about that for the KJMA kinetics, $n = 4$. On the whole, the picture is the same as in the case of spherical geometry of grains [42].

## 7. Conclusions

1. VF equations for the surface-nucleated transformation of a plate, both nonisothermal and isothermal, are obtained by the CR method. The characteristic parameter $\alpha_s$ determining the TK is introduced. The temporal behavior of the Avrami exponent is qualitatively different at small and large values of $\alpha_s$, similarly to the case of a spherical particle.The TK of an ensemble of width-distributed plates slows down relatively the TK of an ensemble of identical plates due to the presence of plates with a large width.
2. The mean-field approximation is used to get a VF equation for the system of an infinite number of parallel planes, both regularly and randomly arranged. For the system of randomly arranged planes with the Poisson point process of their positions, an exact solution is obtained by the CR method; comparison with the MFA solution shows a high accuracy of the latter in this case. A non-Poissonian process of planes positions is considered for the first time. The exact solution obtained for the second-order Erlang process shows that it cannot be derived by the Cahn method, which means that the extended-volume approach is not universal and only applies to Poisson processes.
3. The results obtained for random planes are used to get the VF and grain size distribution for the grain structure of random parallelepipeds formed by three sets of random planes orthogonal to each other, which is a special Cahn model. Based on this (a)-structure, various derivative ones are constructed to the completely uncorrelated (d)-structure which is a random packing of the same parallelepipeds; the latter may approximate a real grain structure. The TK in the (a)-structure is considerably slowed down as compared to that in the (d)-structure. Considering the TK in various intermediate structures between these two, the following rule is established: the TK accelerates, when the degree of correlation in the arrangement of parallelepipeds decreases. In this way, despite the fact that all the structures consist of the same parallelepipeds transforming in the same way, the rates of transformation are different for different packings. This is the paradox of packing suggesting that the correlation in the arrangement of grains in the general Cahn model can be responsible for the TK underestimation as applied to real grain structures. The exponential 1D-growth asymptotics of Cahn’s expression originates just from random planes and, therefore, is closely related to this correlation.

4. VF quations obtained for regularly arranged planes are used to get VF equations for cubic structures, both regular and random. Four different size distributions are employed to study the TK in a random cubic structure. The corresponding Cahn equation (with the same boundaries area in unit volume) underestimates the TK in both regular and random structures for all the distributions considered. The degree of underestimation is different for different size distributions; it is the largest in a regular structure. These results suggest that the correlation effect in the Cahn model acts systematically to underesrimate the TK, but the degree of underestimation also depends on the grain size distribution of the structure under study.

## Appendix

The problem is to find the probability $P_n(x)$ of $n$ points on the segment $[0, x)$, Fig. 14a, in the $E_2$-process. Let the first point be at $x_1 = \Delta_+$, the second point be at $x_2 = \Delta_+ + L_1$, etc., the $n$th point be at $x_n = \Delta_+ + L_1 + L_2 + ... L_{n-1}$, where { $L_k$, $k$ = 1, 2, 3, …} are independent quantities with the probability density and distribution function

$$f(L) = \lambda^2 L \mathrm{e}^{-\lambda L}, \quad F(L) = \int_0^L f(L')dL' = 1 - \mathrm{e}^{-\lambda L}(1 + \lambda L) \tag{A1}$$

(the functions are designated without sub- and superscripts for brevity), whereas the quantity $\Delta_+$ in our case has a different probability density $f_+(L)$. The quantities $\Delta_+$ and $\Delta_-$ correspond to the forward and reverse *return times* in the renewal theory [51]; the probability densities $f_+(L)$ and $f_-(L)$ for these quantities are the same and have the following form [51]:

$$f_+(L) = f_-(L) = \frac{1 - F(L)}{\bar{L}} \tag{A2}$$

which gives

$$f_+(L) = \frac{1}{2}\left(\lambda \mathrm{e}^{-\lambda L} + \lambda^2 L \mathrm{e}^{-\lambda L}\right) \tag{A3}$$

Introducing the distribution function $F_{x_n}(x) = P\{x_n < x\}$ of the coordinates $x_n$, the desired probability $P_n(x)$ is determined as follows [50]:

$$P_n(x) = F_{x_n}(x) - F_{x_{n+1}}(x), \quad n = 1, 2, 3, \dots \tag{A4}$$

$$P_0(x) = 1 - F_{x_1}(x) \tag{A5}$$

As $x_n$ is the sum of $n$ random quantities, the Laplace transformation [51] is convenient to find $F_{x_n}(x)$. The Laplace transformations of the probability density $f(L)$ and the distribution function $F(L)$ of a non-negative random quantity $L$ are

$$f^*(s) = \overline{\exp(-sL)} = \int_0^\infty e^{-sL} f(L)dL, \quad F^*(s) = f^*(s)/s \tag{A6}$$

where the bar denotes a mathematical expectation. If $L_1$, $L_2$, …, $L_n$ are non-negative independent random quantities with the probability densities $f_1(L)$, $f_2(L)$, …, $f_n(L)$, then the Laplace transformation $f^*_{x_n}(s)$ of the probability density $f_{x_n}(x)$ of the sum $x_n = L_1 + L_2 + ... + L_n$ by definition is

$$f^*_{x_n}(s) = \overline{\exp(-sx_n)} = \overline{\exp[-s(L_1 + L_2 + ... + L_n)]} = f_1^*(s) f_2^*(s) ... f_n^*(s) \tag{A7}$$

In particular, if $f_1(L) = f_2(L) = \ldots = f_n(L) = f(L)$, then

$$f^*_{x_n}(s) = \left[f^*(s)\right]^n \tag{A8}$$

In this way, we have for the given problem

$$F^*_{x_n}(s) = \frac{f_+^*(s)[f^*(s)]^{n-1}}{s} \tag{A9}$$

with

$$f^*(s) = \frac{\lambda^2}{(s+\lambda)^2}, \quad f_+^*(s) = \frac{1}{2}\left(\frac{\lambda}{s+\lambda} + \frac{\lambda^2}{(s+\lambda)^2}\right) \tag{A10}$$

i. e.

$$F^*_{x_n}(s) = \frac{1}{2s}\left[\left(\frac{\lambda}{s+\lambda}\right)^{2n-1} + \left(\frac{\lambda}{s+\lambda}\right)^{2n}\right] \tag{A11}$$

Expansion into simple fractions has the following form:

$$F^*_{x_n}(s) = \frac{1}{s} - \frac{1}{\lambda}\sum_{i=1}^{2n}\frac{\lambda^i}{(s+\lambda)^i} + \frac{1}{2\lambda}\frac{\lambda^{2n}}{(s+\lambda)^{2n}} \tag{A12}$$

Further, the desired original $F_{x_n}(x)$ for $F^*_{x_n}(s)$ is found with the use of originals $x^{i-1} e^{-\lambda x}/(i-1)!$ for $1/(s+\lambda)^i$ and 1 for $1/s$:

$$F_{x_n}(x) = 1 - e^{-\lambda x}\left\{\sum_{i=0}^{2n-1}\frac{(\lambda x)^i}{i!} - \frac{1}{2}\frac{(\lambda x)^{2n-1}}{(2n-1)!}\right\} \tag{A13}$$

Finally, the application of Eqs. (A4) and (A5) yields

$$P_n(x) = e^{-\lambda x}\left\{\frac{1}{2}\frac{(\lambda x)^{2n+1}}{(2n+1)!} + \frac{(\lambda x)^{2n}}{(2n)!} + \frac{1}{2}\frac{(\lambda x)^{2n-1}}{(2n-1)!}\right\}, \quad n = 1, 2, 3, \ldots \tag{A14}$$

$$P_0(x) = \mathrm{e}^{-\lambda x}\left(1 + \frac{1}{2}\lambda x\right) \quad \text{(A15)}$$

It is seen that the probability $P_n(x)$ is the sum of three Poisson probabilities: one for even ($2n$) number of points and two for odd numbers ($2n-1$) and ($2n+1$); the latter enter with the weighting factor 1/2. Such expression structure reflects the fact that $n$ points on the segment $[0,x)$ in the given E$_2$-process can be obtained as result of thinning out the parent P-process containing on the segment $[0,x)$ either $2n-1$, $2n$, or $2n+1$ points. These possibilities are depicted in Fig. 14b by the example of $n=2$.

To check the fulfillment of the full probability condition, we note that

$$\sum_{n=1}^{\infty} P_n(x) = \mathrm{e}^{-\lambda x}\left\{\sum_{m=1}^{\infty}\frac{(\lambda x)^m}{m!} - \frac{1}{2}\lambda x\right\} = 1 - \mathrm{e}^{-\lambda x}\left(1 + \frac{1}{2}\lambda x\right) = 1 - P_0(x) \quad \text{(A16)}$$

i. e. the full probability equals to unity, as it must.

It is also of interest to find the mean value of $n(x)$, $H(x) = \bar{n}(x)$, which is called the *renewal function* [51]. By definition and with the use of Eq. (A4),

$$H(x) = \sum_{n=0}^{\infty} nP_n(x) = \sum_{n=0}^{\infty} n[F_{x_n}(x) - F_{x_{n+1}}(x)] = \sum_{n=1}^{\infty} F_{x_n}(x) \quad \text{(A17)}$$

Taking the Laplace transformation of this equation and using Eqs. (A9) and (A10), we get

$$H^*(s) = \frac{f_+^*(s)}{s(1 - f^*(s))} = \frac{\lambda(s+\lambda) + \lambda^2}{2s^2(s+2\lambda)} = \frac{\lambda}{2}\frac{1}{s^2} \quad \text{(A18)}$$

which corresponds to

$$H(x) = \bar{n}(x) = \frac{\lambda}{2}x \quad \text{(A19)}$$

As it was expected, the mean $\bar{n}(x)$ is half the mean $\lambda x$ of the parent P-process.

In conclusion, it should be noted that the function $f_+(L)$, Eq. (A2), is inherent in the so-called *stationary* renewal process [51] which just corresponds to the considered problem. For the exponential distribution, the equality $f_+(L) = f(L) = \lambda\mathrm{e}^{-\lambda L}$ holds; being applied to the P-process, the above procedure results in the Poisson distribution $P_{(1)}(n)$, Eq. (39b).

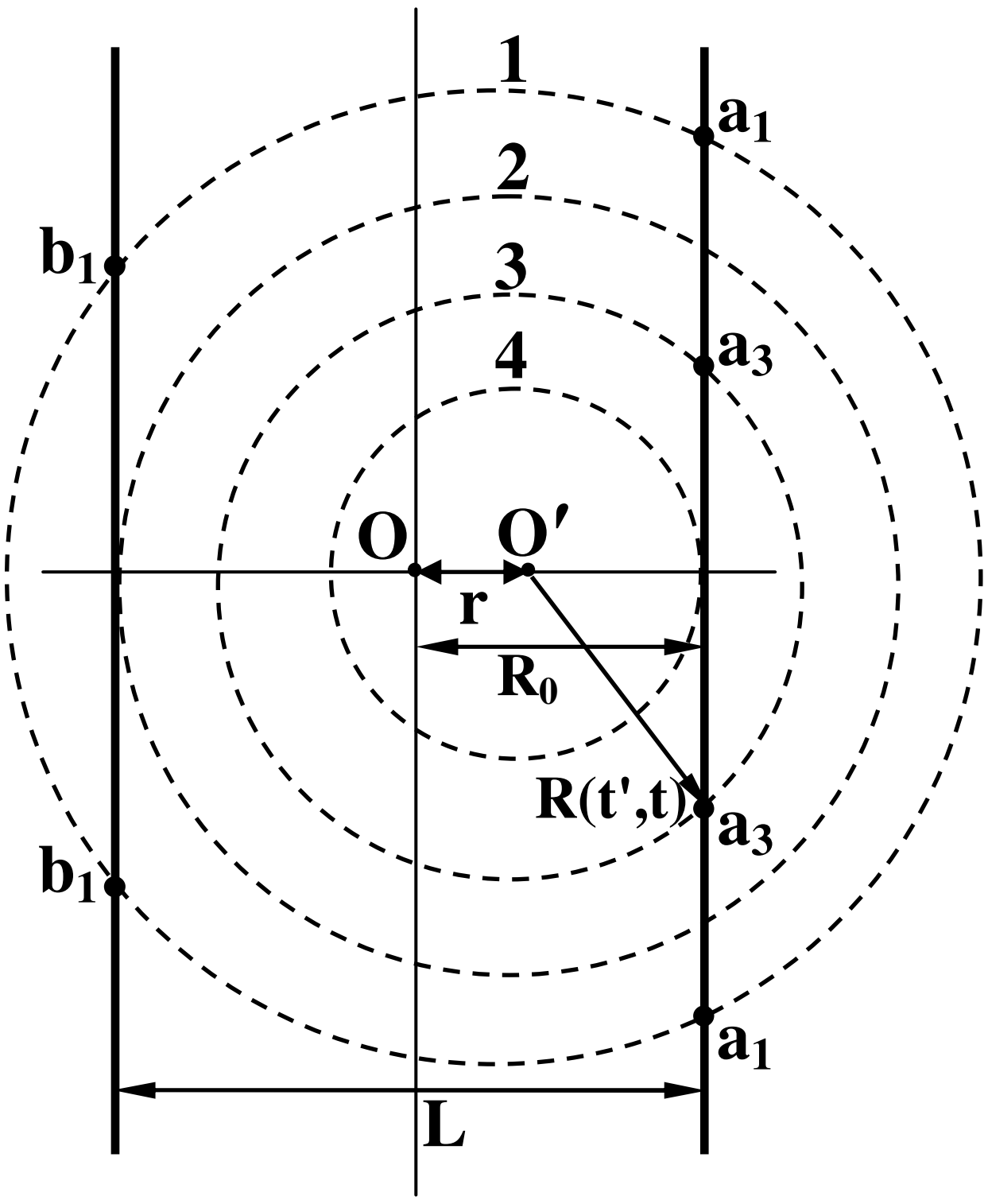

Fig. 1. Positions of the CR boundary at different times $t'$ in cases (2) and (3): (1) $0 \le t' < t'_m(r,t)$; (2) $t' = t'_m(r,t)$; (3) $t'_m(r,t) < t' < t_m(r,t)$; (4) $t' = t_m(r,t)$.

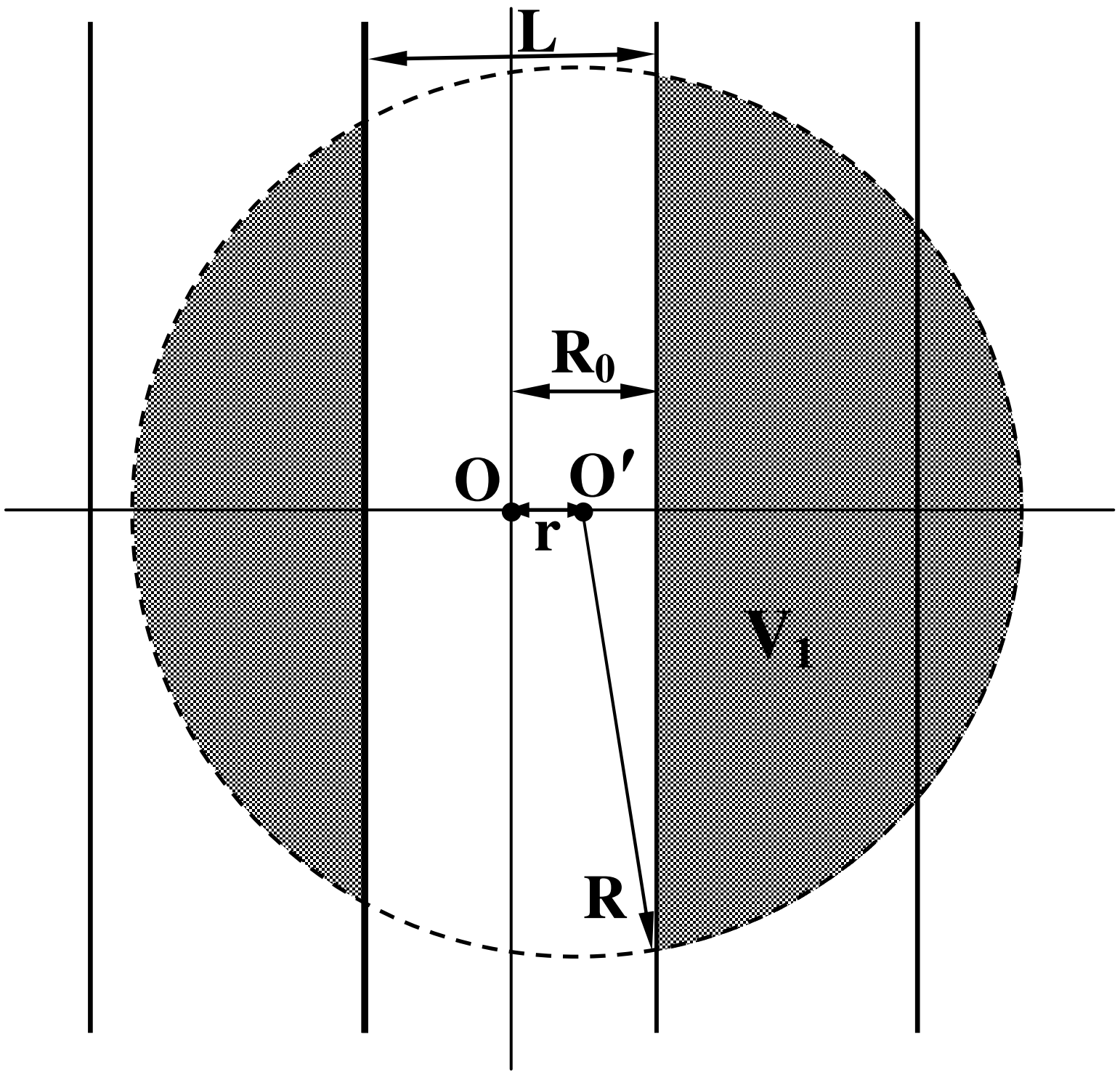

Fig. 4. To calculating the VF in the system of an infinite number of parallel planes in the MFA; explanations are in the text.

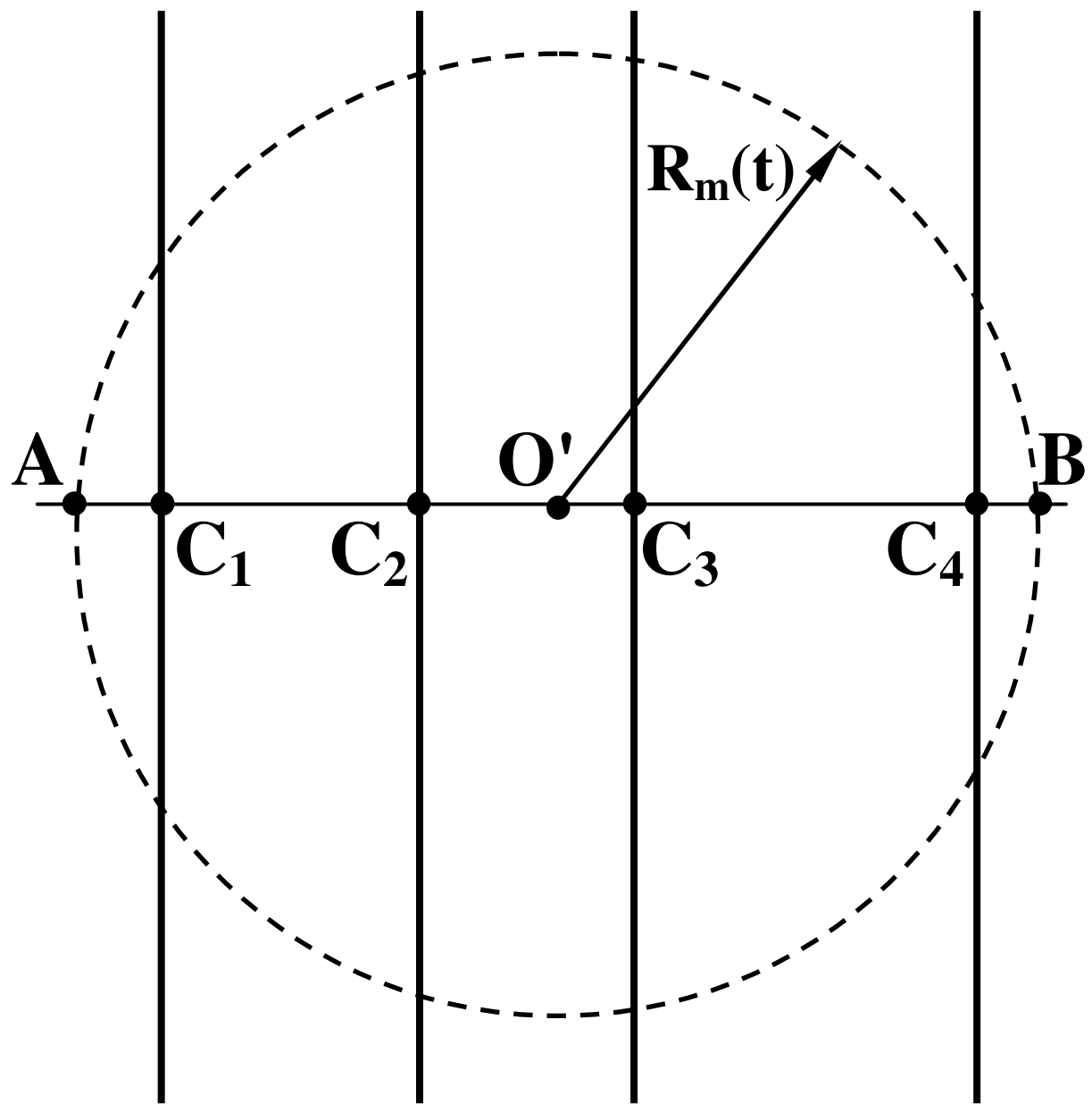

Fig. 6. To exact calculating the VF in the system of an infinite number of random parallel planes; explanations are in the text.

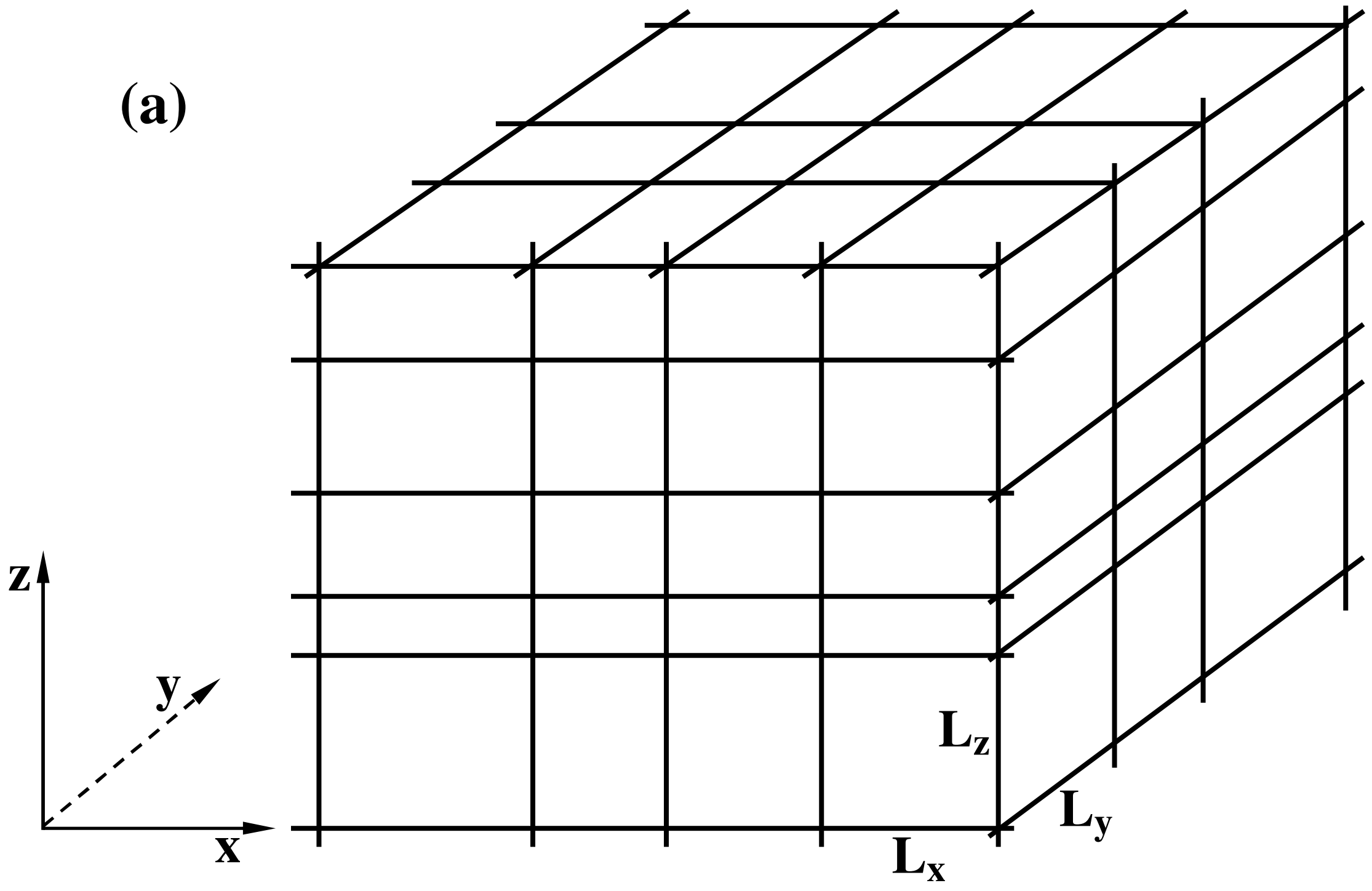

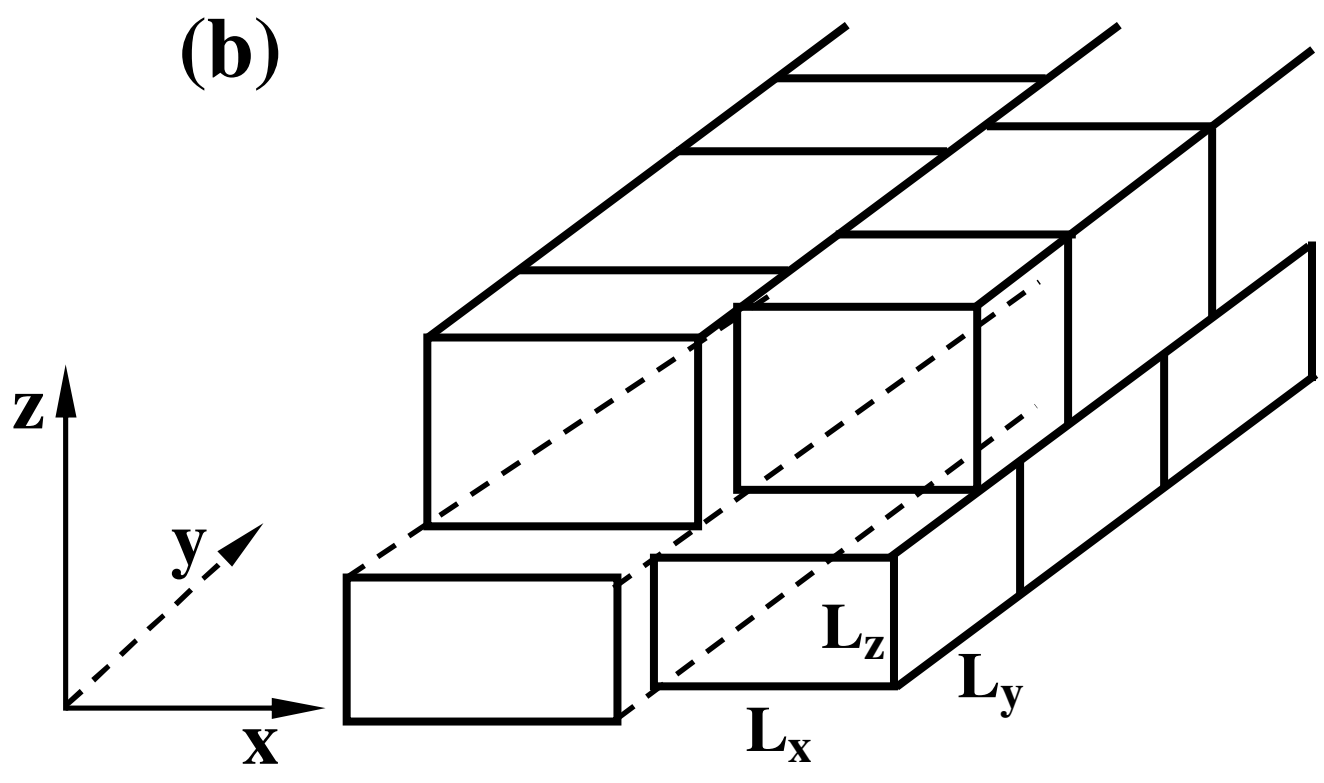

(c')

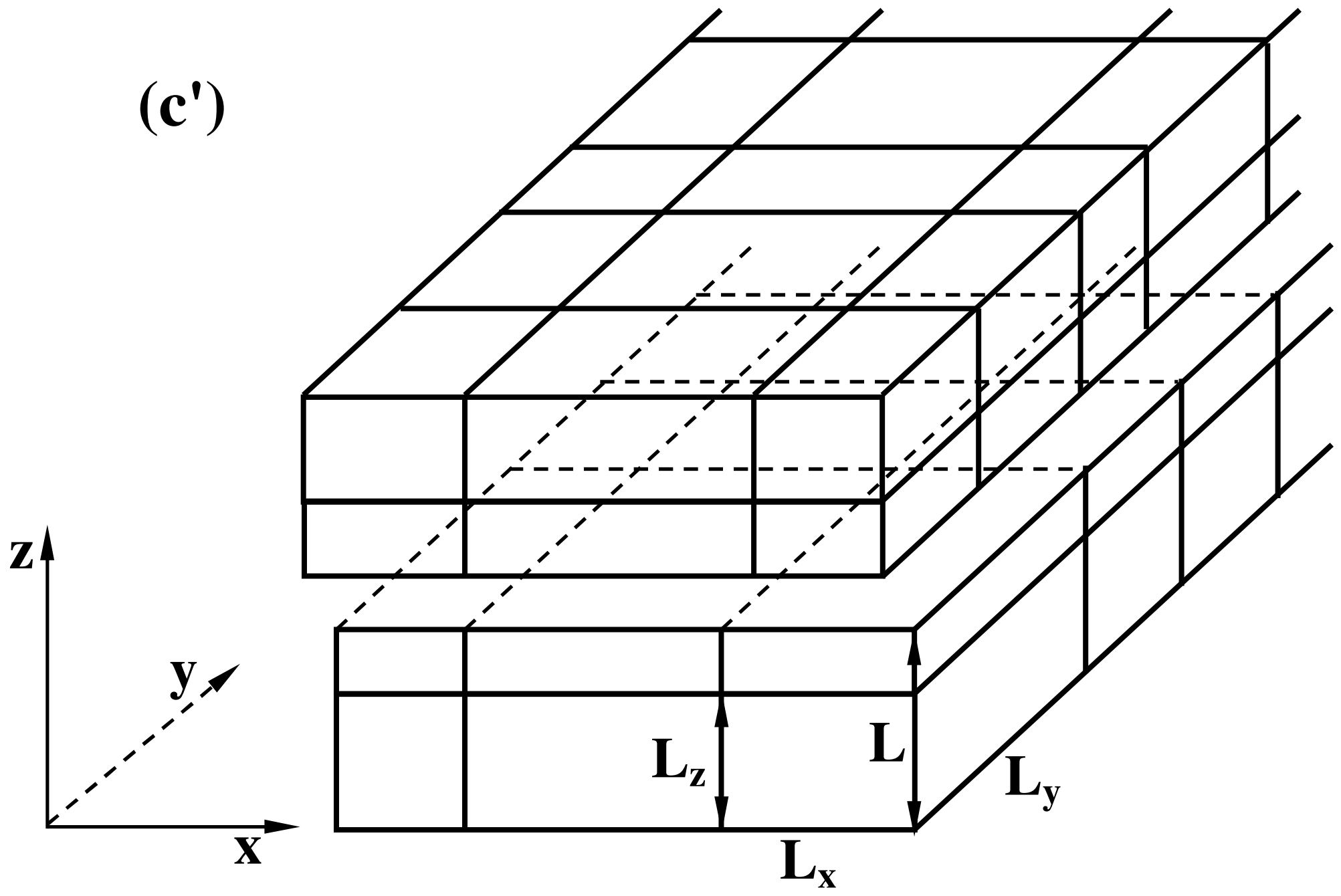

(c)

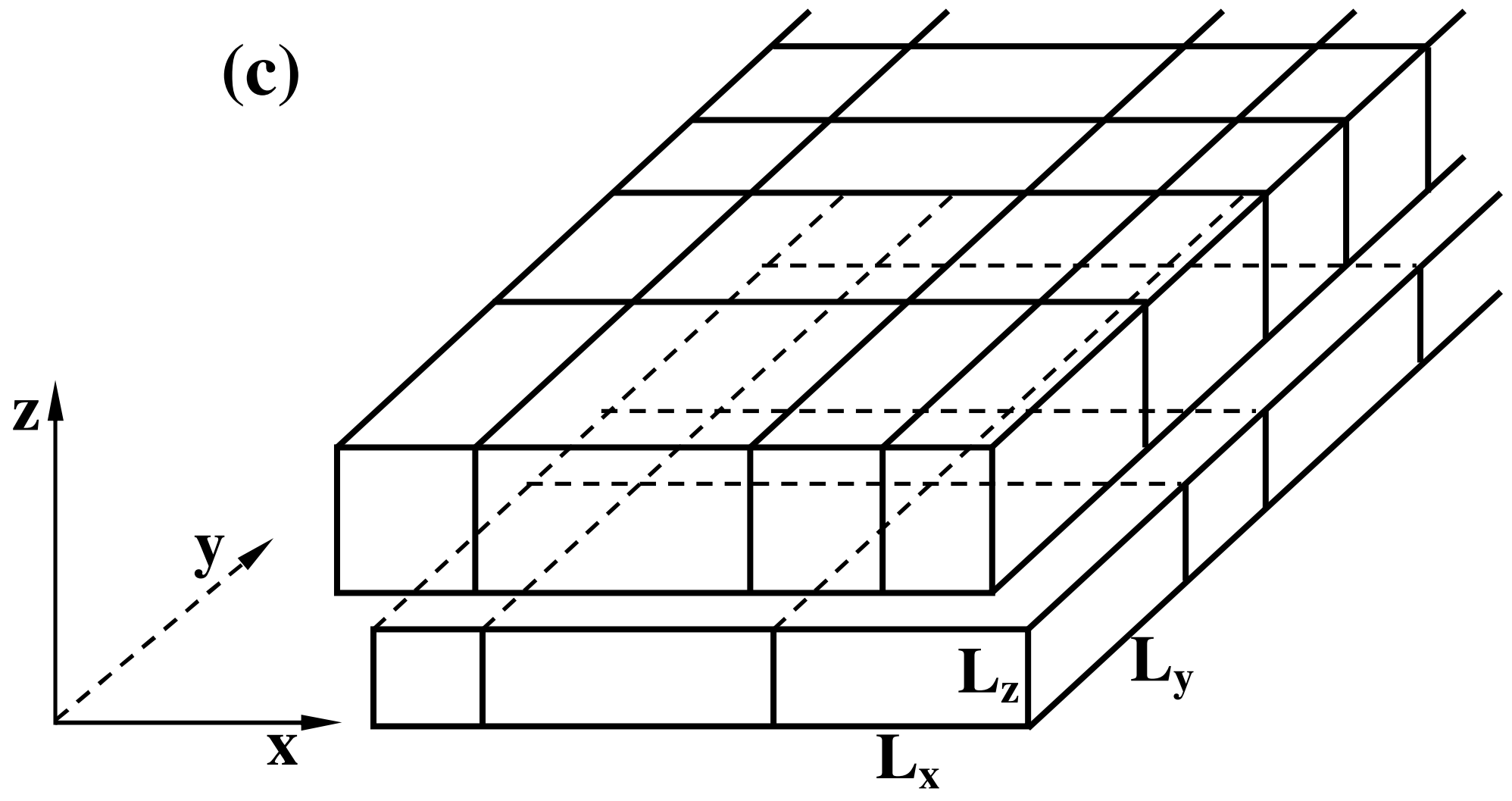

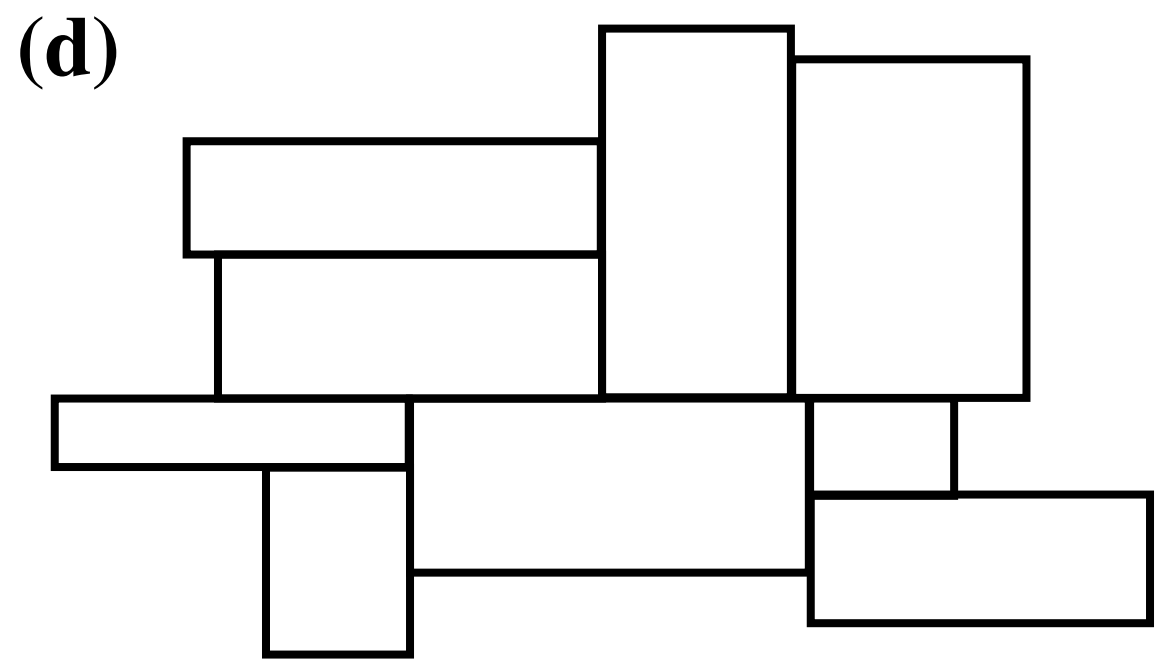

Fig. 8. (a) Grain structure formed by three sets of random parallel planes orthogonal to each other and therefore consisting of random parallelepipeds. The gradual weakening of correlation in the arrangement of parallelepipeds occurs in structures (b) – (d); explanations are in the text. (d) 2D section of the random packing of parallelepipeds from Fig. (a).

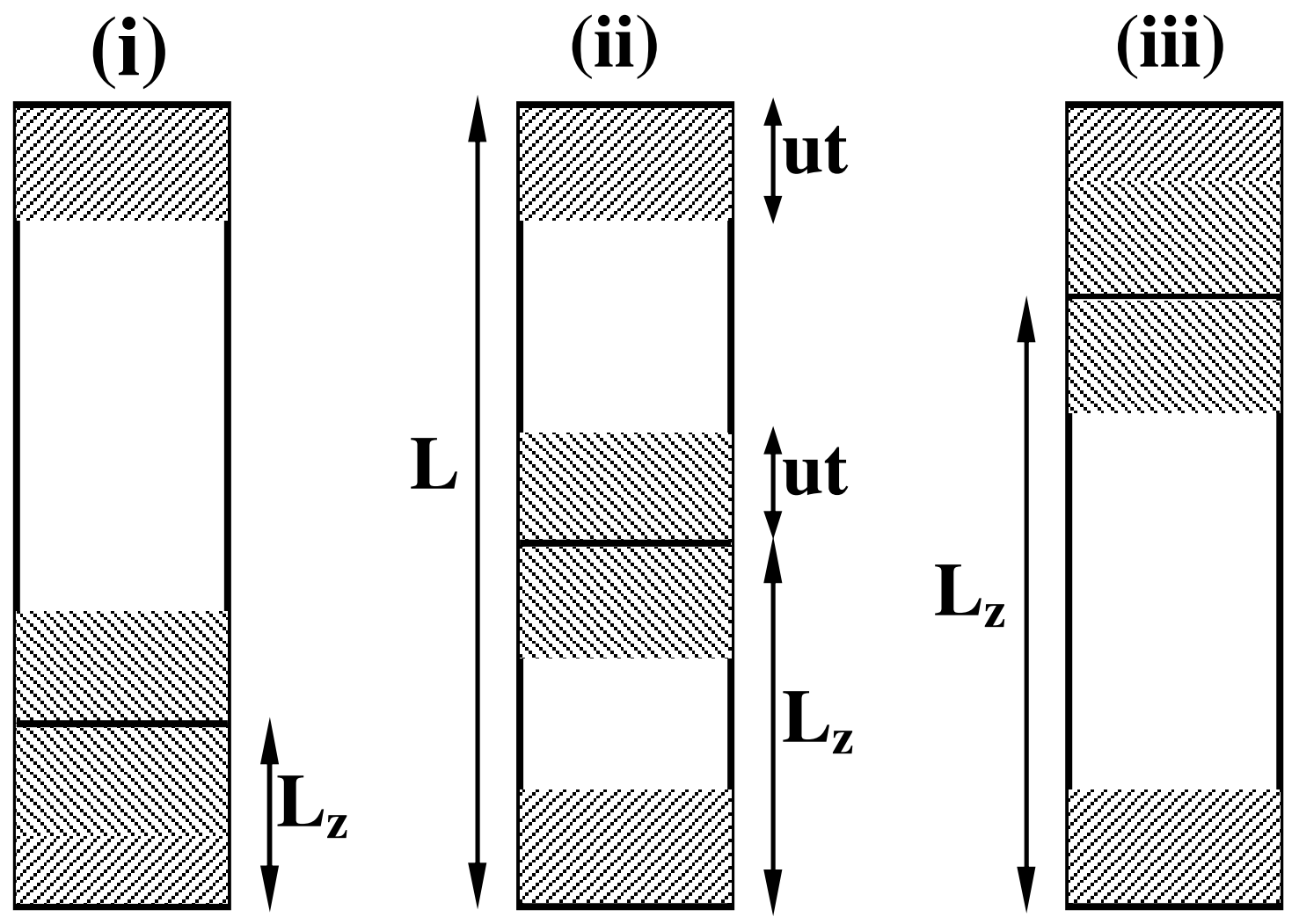

Fig. 10. To calculating the function $c'(\bar{\tau})$. The transformed part of the length $L$ at time $t$ is shaded; other explanations are in the text.

(a)

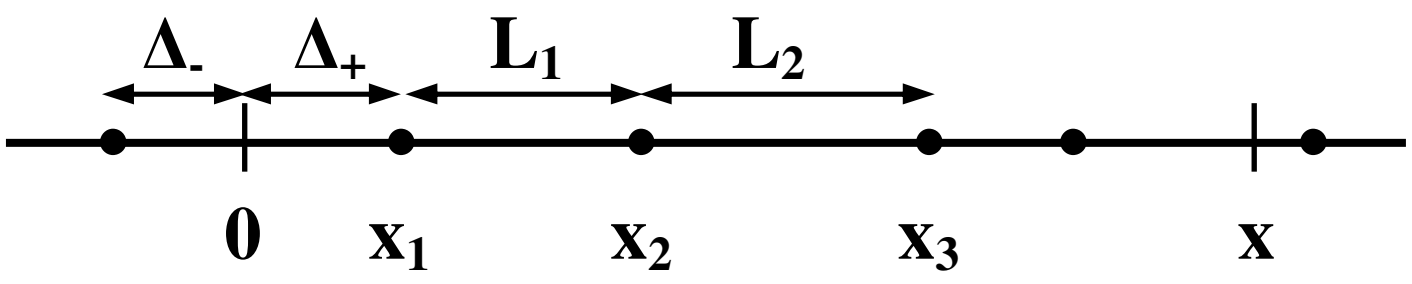

(b)

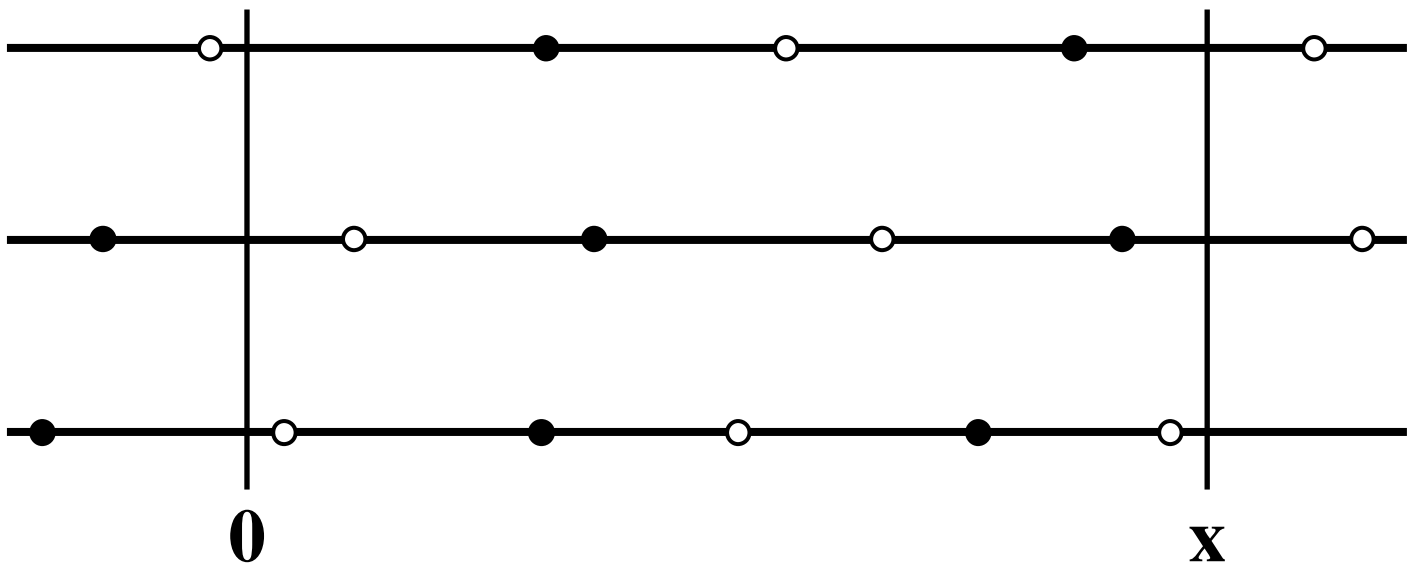

Fig. 14. (a) To calculating the probability $P_n(x)$. (b) Sketch exemplifying Eq. (A14) for $n=2$. White points drop out from the P-process as a result of its thinning out; the two (black) points of the resulting $E_2$-process remain on the segment $[0,x)$ from the initial either three, four, or five points of the parent P-process.

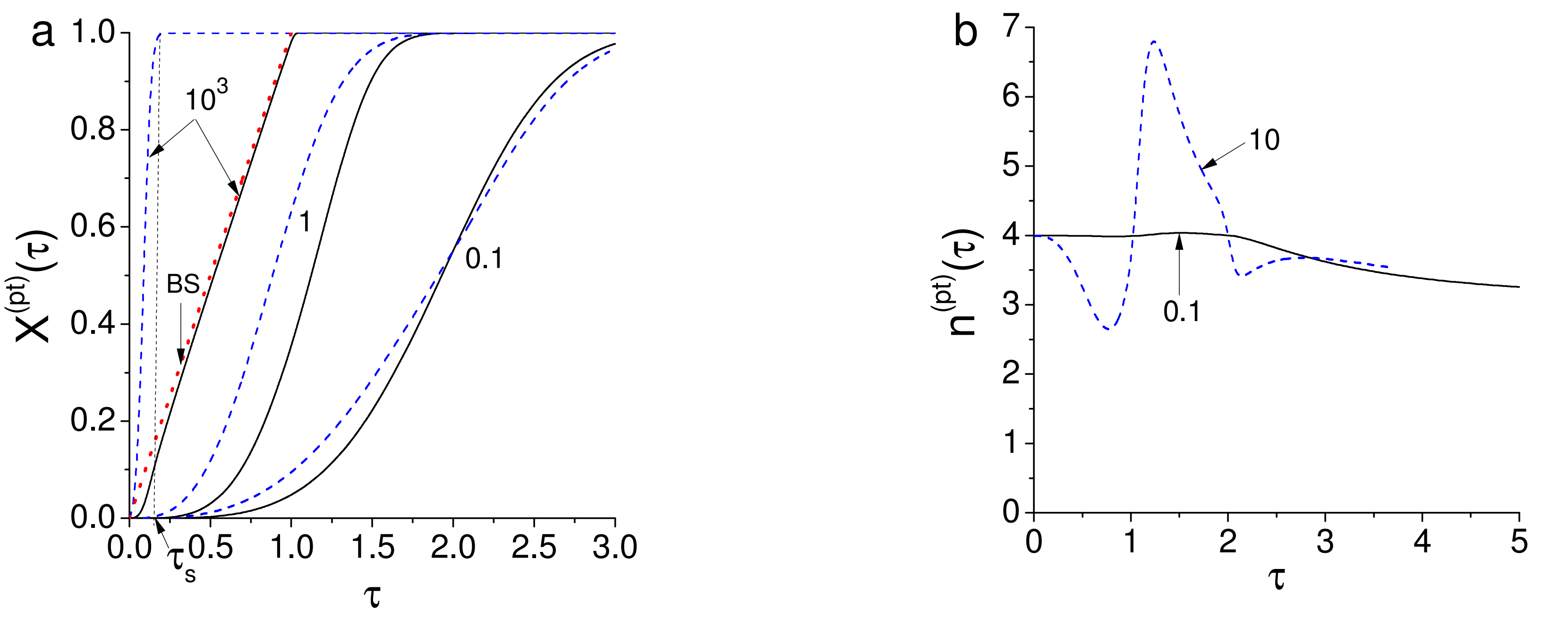

Fig. 2 a, b

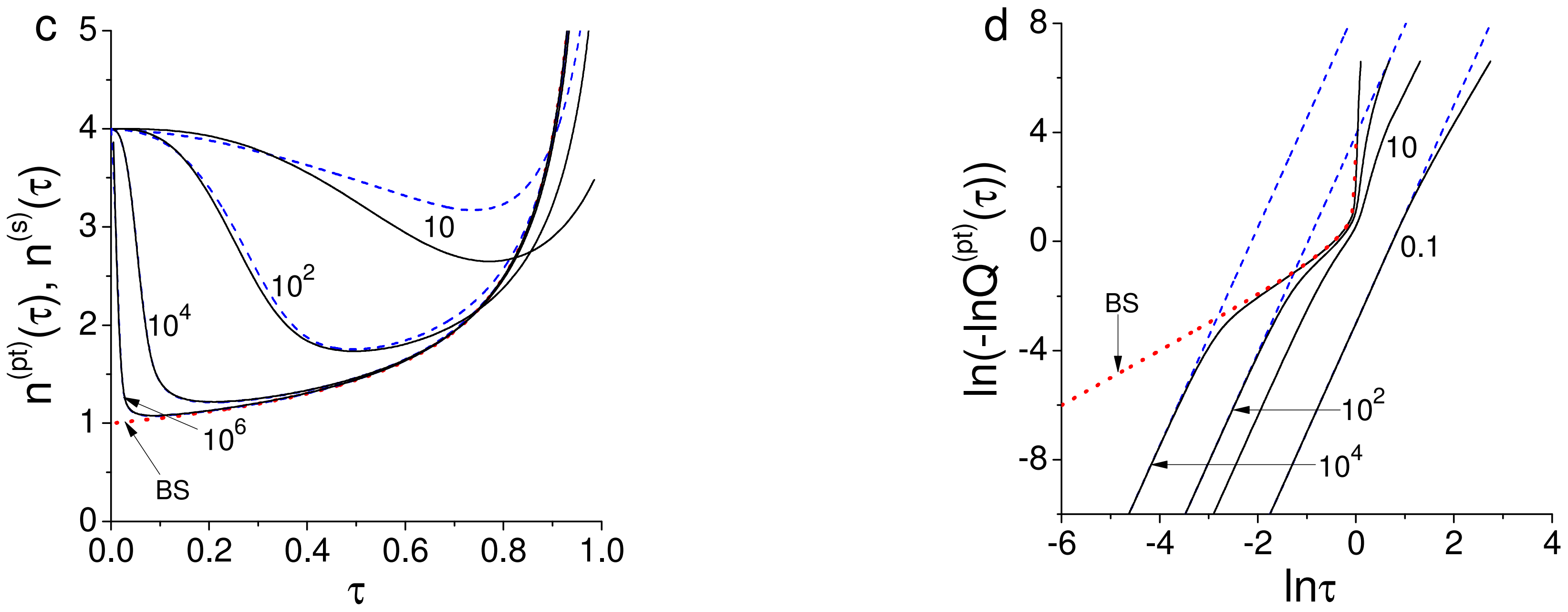

Fig. 2. (a) Kinetic curves for the transformation of a plate (solid) and its surface (dashed) at different values of the characteristic parameter $\alpha_s$ shown at the curves. The BS-limit line $X_{BS}^{(pt)}(\tau) = \tau$ is dotted; $\tau_s$ is the BS time, when the surface transformation is completed. (b) Full-range temporal behavior of the Avrami exponent for two $\alpha_s$ values. (c) Temporal behavior of the Avrami exponent for the plate (solid) and a spherical particle [42] (dashed) at the same (large) $\alpha_s$ values; the BS-limit curve (dotted) is given by Eq. (14). (d) Double-logarithmic VF plots for the plate (solid) together with the corresponding KJMA lines for bulk nucleation (dashed) at different $\alpha_s$ values shown at the curves.

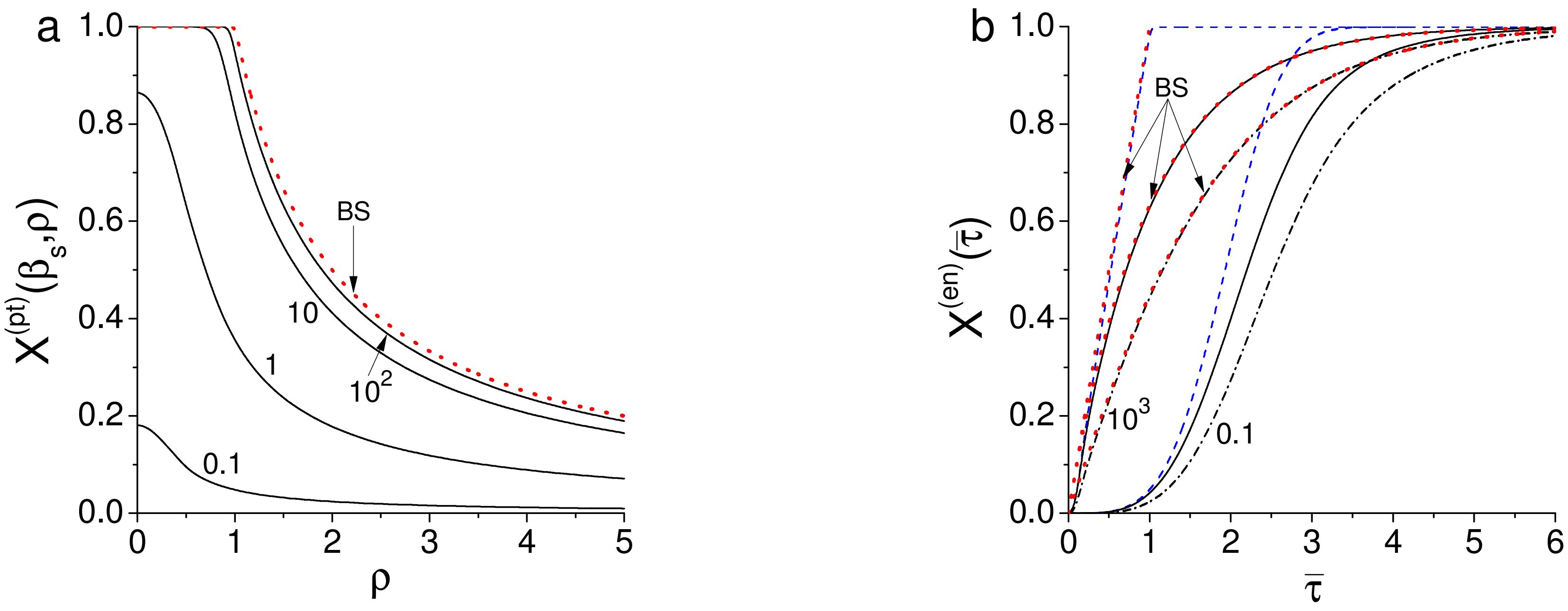

Fig. 3. (a) Radius-dependent VF $X^{(pt)}(\beta_s, \rho)$ of a plate for different $\beta_s$ values shown at the curves. (b) Kinetic curves for the ensemble of width-distributed plates with the exponential (solid) and second-order Erlang (dash-dotted) distributions versus the curves for the ensemble of identical plates (dashed) with $\alpha_s = \bar{\alpha}_s$. At $\alpha_s = 10^3$, each curve is close to its own BS-limit line, $X_{BS}^{(pt)}(\tau) = \tau$ and Eqs. (26a, b).

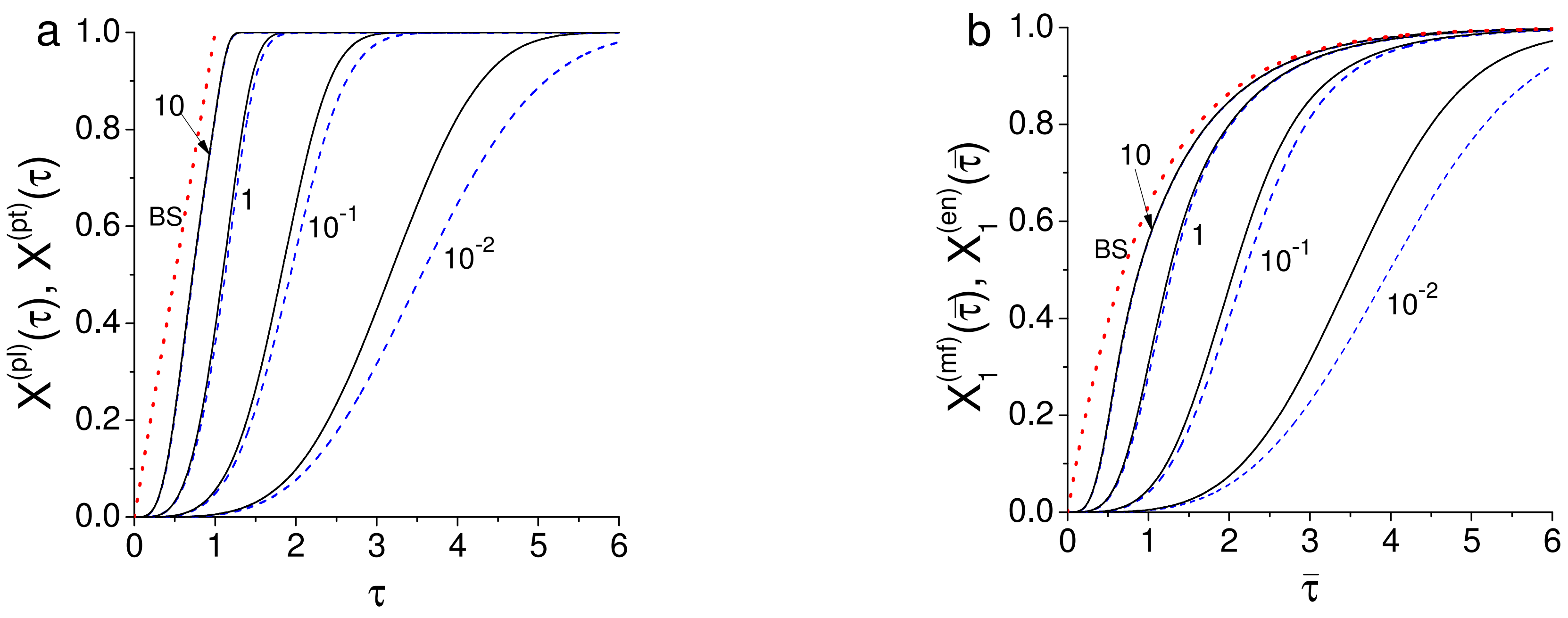

Fig. 5. MFA kinetic curves $X^{(pl)}(\tau)$ for the ensemble of regularly arranged planes (solid) versus the kinetic curves $X^{(pt)}(\tau)$ for the ensemble of isolated identical plates (dashed) at different $\alpha_s$ values. (b) MFA kinetic curves $X_1^{(mf)}(\bar{\tau})$ for the ensemble of randomly arranged planes with the exponential distribution of spacing between them (solid) versus the kinetic curves $X_1^{(en)}(\bar{\tau})$ for the ensemble of isolated width-distributed plates with the exponential distribution (dashed) at different $\alpha_s$ values.

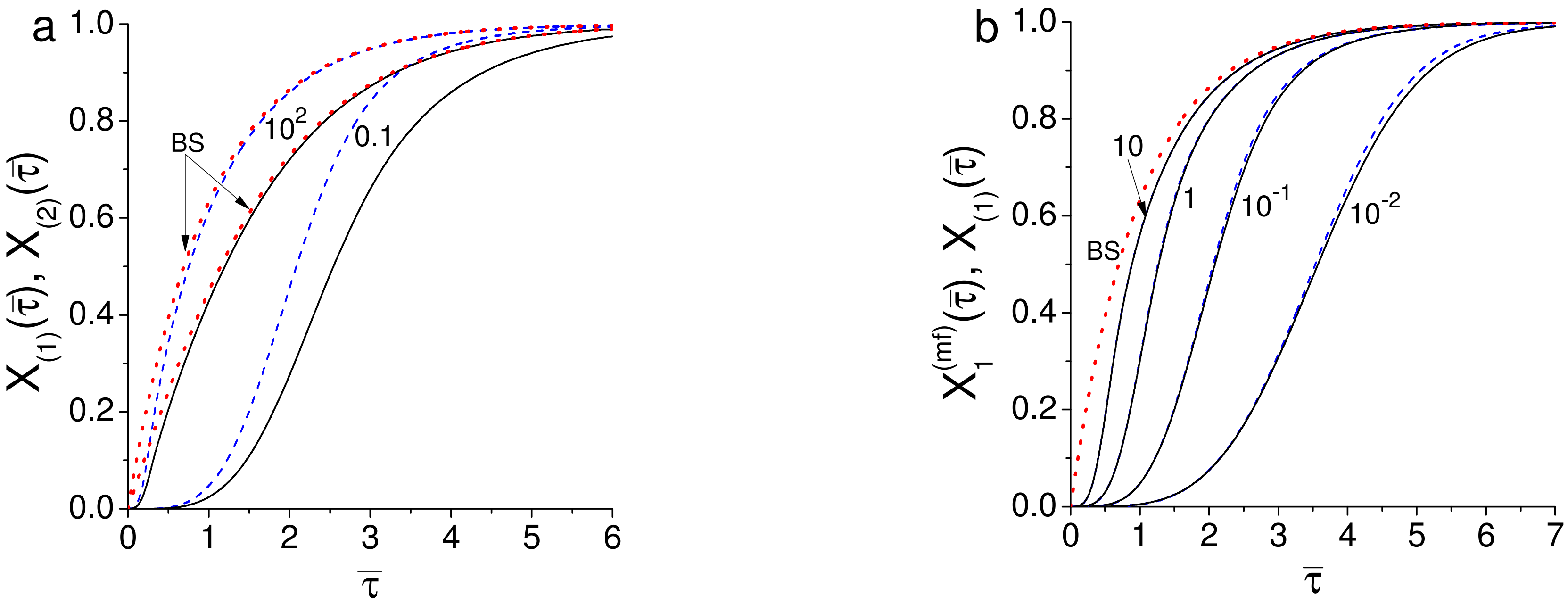

Fig. 7a, b.

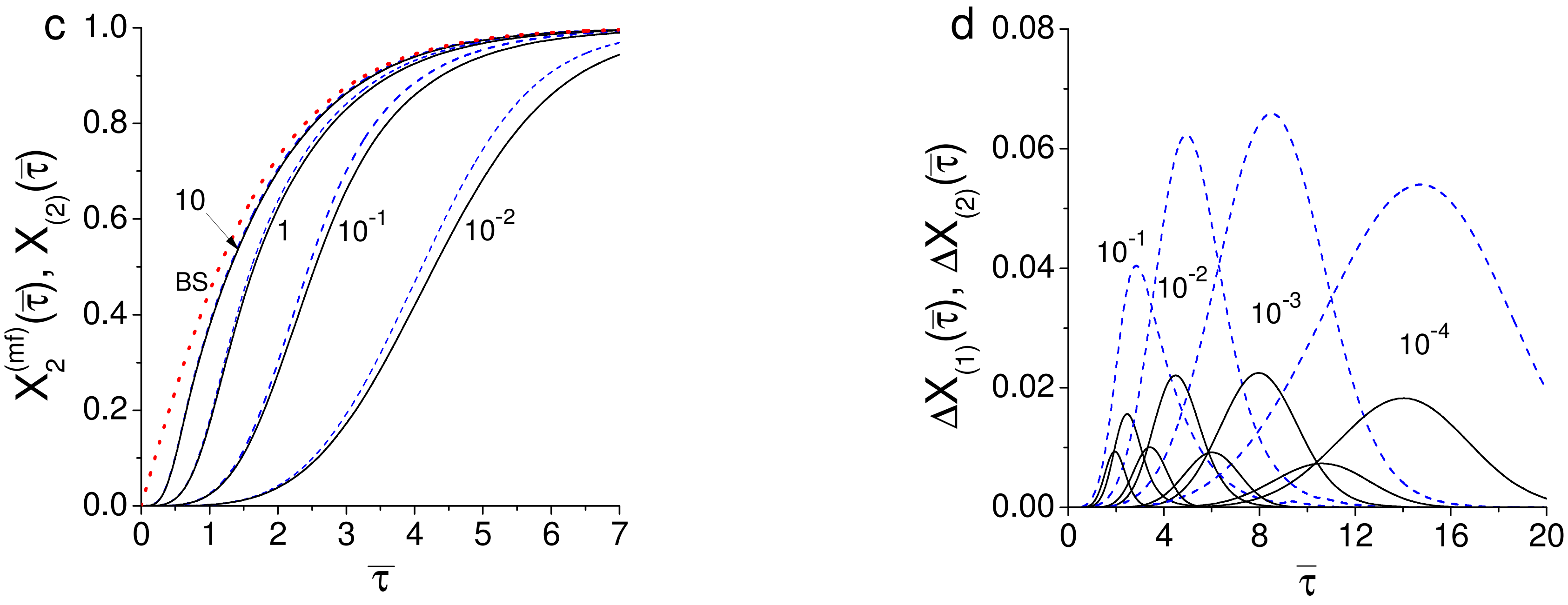

Fig. 7. (a) Kinetic curves $X_{(1)}(\bar{\tau})$ (dashed) and $X_{(2)}(\bar{\tau})$ (solid) for the Poisson and second-order Erlang distribution of planes, respectively, as a result of exact solution. The corresponding BS-limit lines (dotted) are given by Eqs. (26a, b). (b) MFA (dashed) versus exact (solid) kinetic curves for the P-process of planes at different $\alpha_s$ values. (c) MFA (dashed) versus exact (solid) kinetic curves for the $E_2$-process of planes. (d) MFA error for the P-process (top solid) and $E_2$-process (dashed) at different $\alpha_s$ values, as well as MFA error for the VF cubes $\Delta X_{(1)}^{(c)}(\bar{\tau})$ (bottom solid).

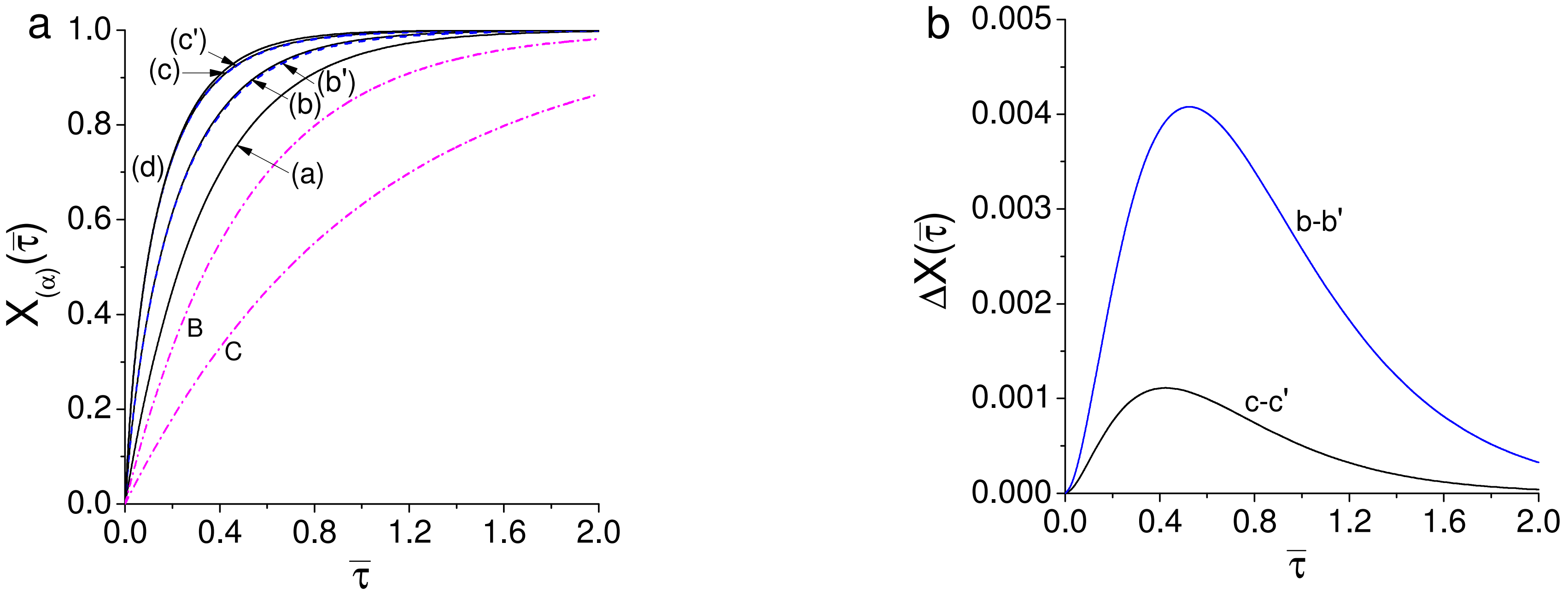

Fig. 9. (a) Kinetic curves $X_{(\alpha)}(\bar{\tau})$, $\alpha =$ a, b, c, d (solid), b', c' (dashed), and B, C (dash-dotted) for the structures of Fig. 8. (b) Differences between curves (c) and (c') as well as (b) and (b').

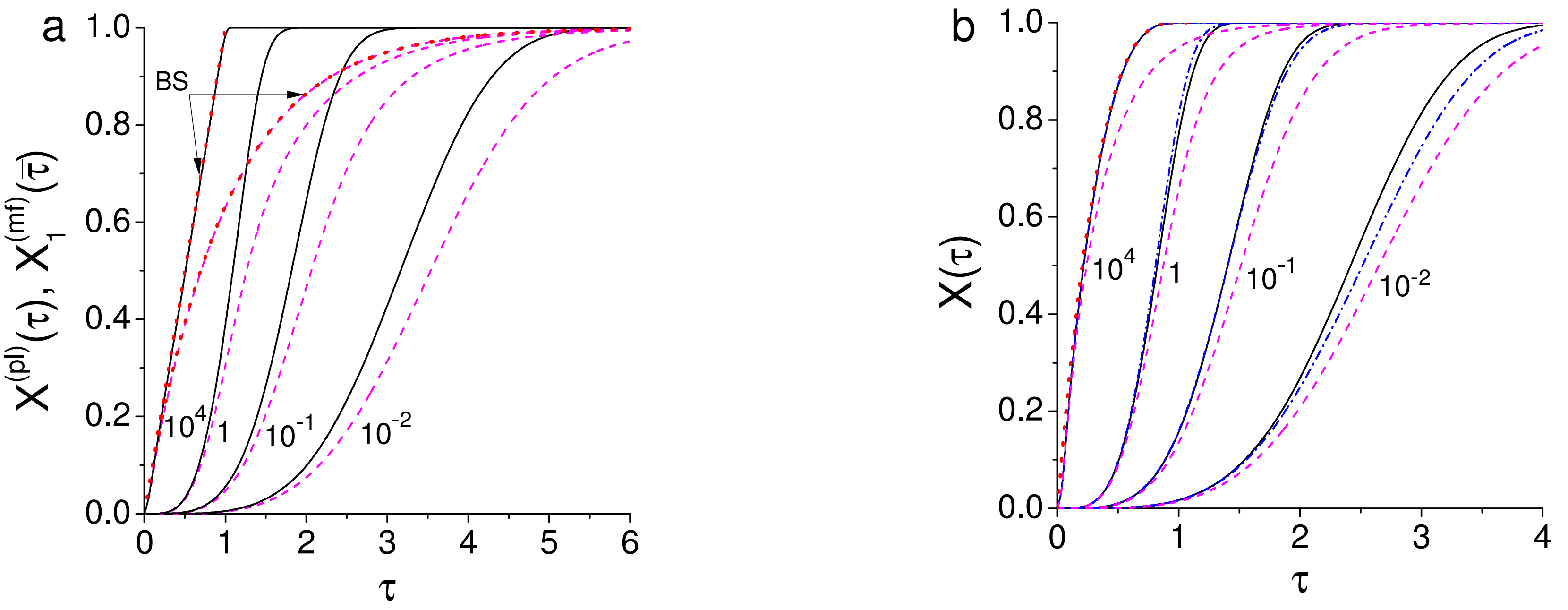

Fig. 10. (a) Kinetic curves for regular (solid) and random (dashed) planes with the exponential distribution (the 1D Cahn model) for different $\alpha_s$ values. (b) Kinetic curves for the regular cubic structure (solid), the dual structure of inscribed spheres (dash-dotted), and Cahn's model (dashed) for different $\alpha_s$ values. The BS-limit line (dotted) represents the equation $X_{BS}^{(c)}(\tau) = 1 - Q_{BS}^{(c)}(\tau)$, Eq. (77).

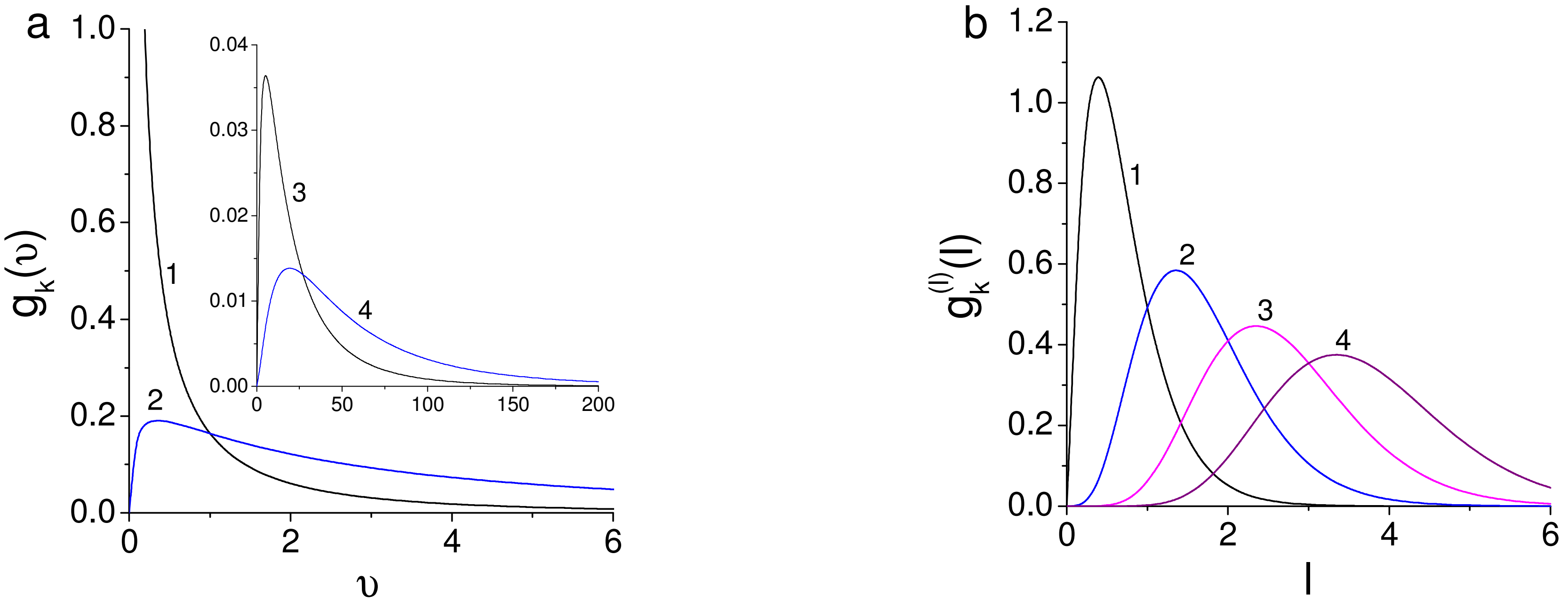

Fig. 11. (a) Volume distribution $g_k(\upsilon)$ of parallelepipeds in the (a)-structure for $k = 1$, 2, 3, and 4, Eq. (57b). (b) Size distribution of cubes of the same volumes, Eq. (57c).

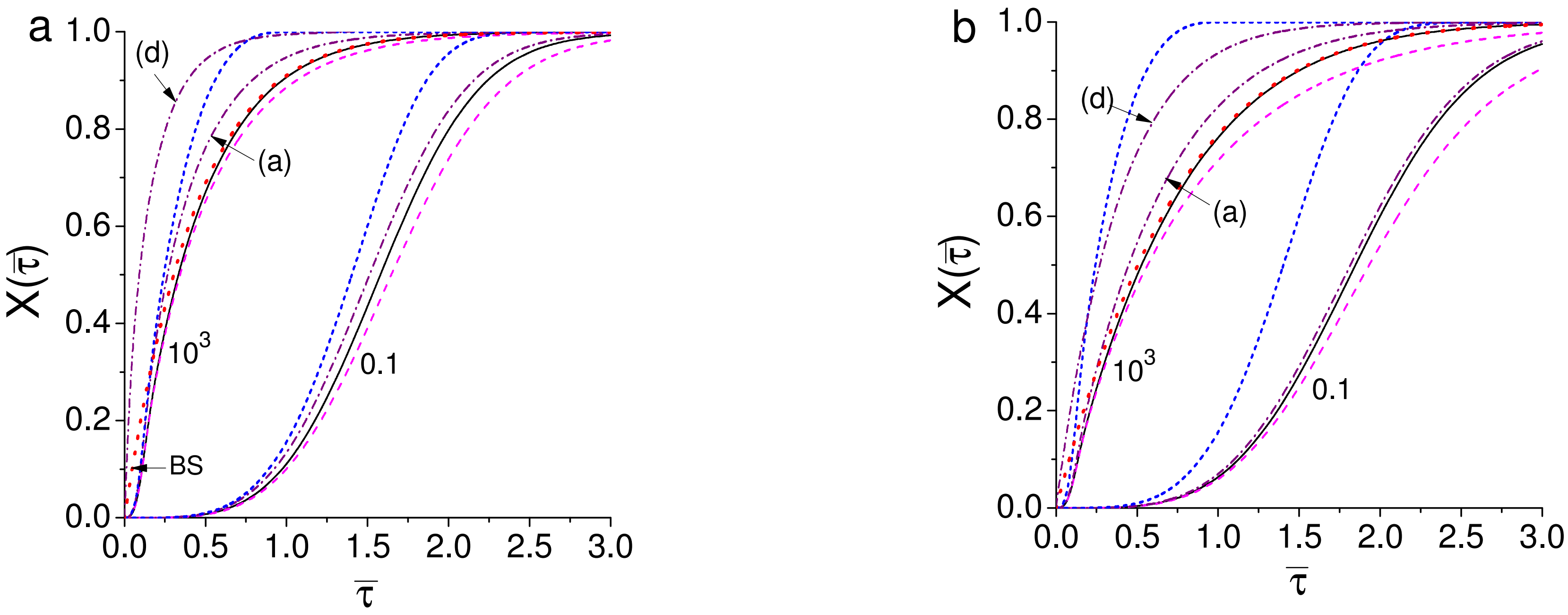

Fig. 12a, b.

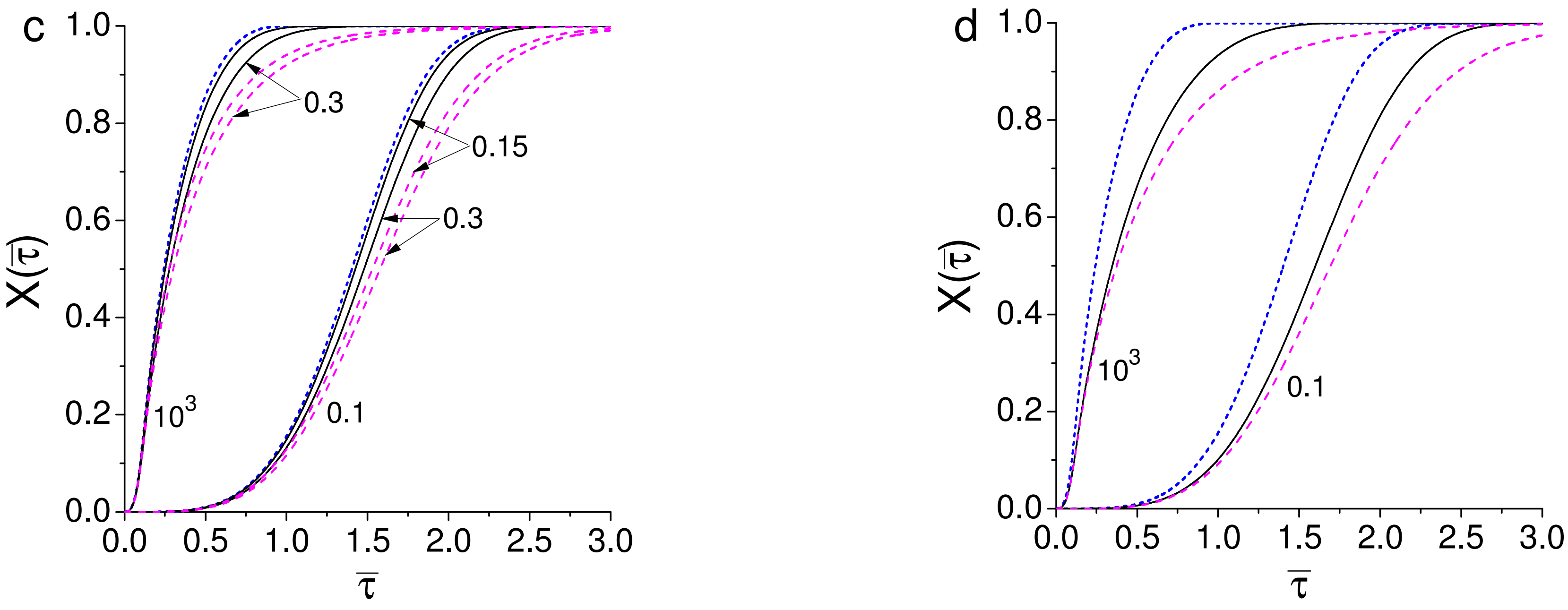

Fig. 12. Kinetic curves for the structure of size-distributed cubes (solid) for the $g_1^{(l)}(l)$ (a), $g_2^{(l)}(l)$ (b), normal with $\sigma = 0.15$ and 0.3 (c), and uniform (d) distributions. Other curves are for the following structures: regular cubic (short dash), Cahn's model (dashed), the (a)- and (d)-structures of parallelepipeds of Fig. 8 (dash-dotted); the BS-limit line for size-distributed cubes, Eq. (81), is dotted.

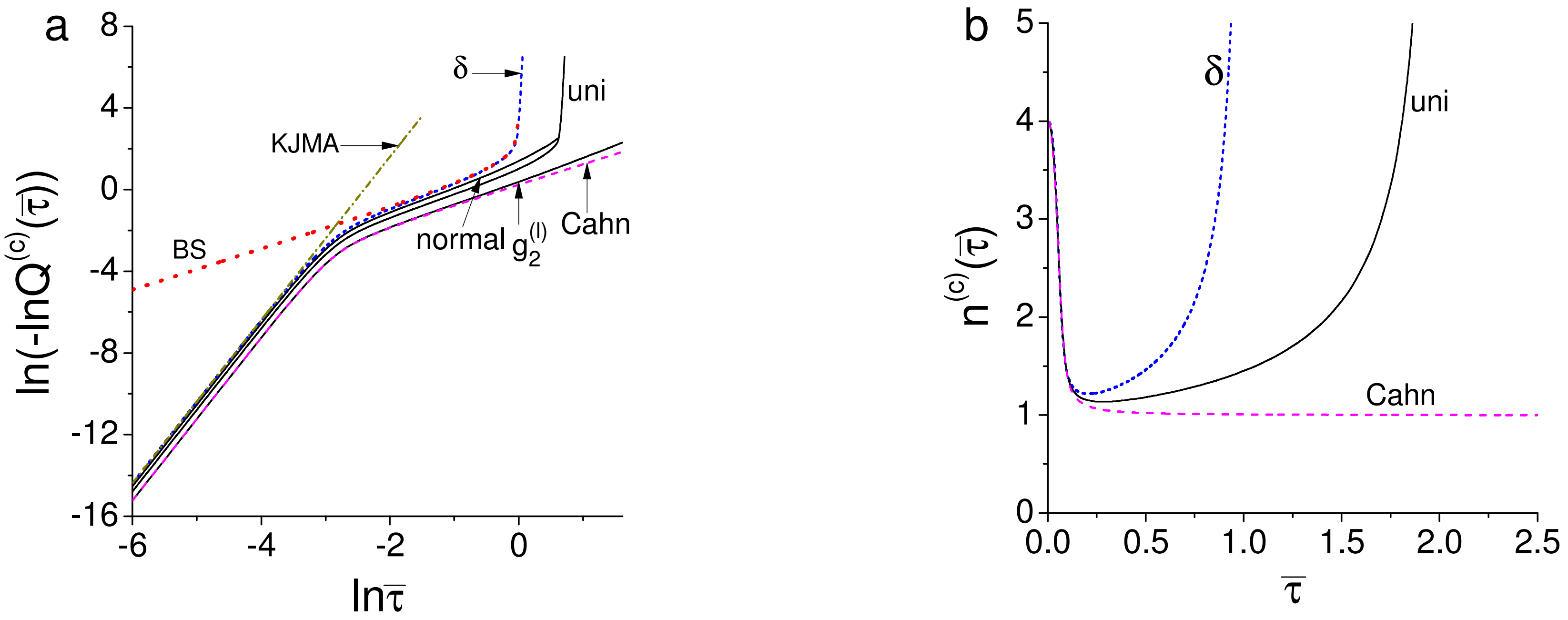

Fig. 13. Double-logarithmic VF curves for the grain structure of size-distributed cubes for different distributions (a) and the corresponding temporal behavior of the Avrami exponent for the $\delta$-shaped and uniform distributions (b).